\documentclass{jfm}

\usepackage{graphicx}
\usepackage[dvipsnames]{xcolor}
\usepackage{newtxtext}
\usepackage{newtxmath}
\usepackage{natbib}
\usepackage{hyperref}
\usepackage{comment}
\hypersetup{
    colorlinks = true,
    urlcolor   = blue,
    citecolor  = blue
}

\newcommand{\RomanNumeralCaps}[1]
\linenumbers

\title{Integral analysis based diagnostics of turbulence model errors in skin friction}

\author{Shyam S. Nair\aff{1}
  \corresp{\email{shyam.nair@psu.edu}},
   Vishal A. Wadhai\aff{1}, Robert F. Kunz\aff{1} \and Xiang I. A. Yang\aff{1} \corresp{\email{xzy48@psu.edu}}
  }

\affiliation{\aff{1}Mechanical Engineering, Penn State University, State College, PA 16802, USA}

\begin{document}
\maketitle

\begin{abstract}
Error diagnostics for turbulence models have traditionally focused on engineering quantities of interest, such as the skin-friction coefficient, $C_f$, most often by comparing the predicted $C_f$ against reference data. In wall-bounded turbulent boundary layers, however, $C_f$ results from several physical mechanisms --- viscous effects, turbulence, pressure gradients, and mean-flow development---whose relative importance depends on the flow conditions. Modeling errors in these mechanisms vary across turbulence closures, and identifying them offers valuable physical insight for model evaluation and improvement.  
We propose a diagnostics framework that systematically isolates and quantifies such errors using the angular momentum integral (AMI) formulation. The method is applied to five transport-type Reynolds-averaged Navier-Stokes (RANS) models in two test cases: a canonical zero-pressure-gradient flat-plate boundary layer and flow over a three-dimensional hill.  
For the flat-plate case, comparison with direct numerical simulation (DNS) data shows that all models reproduce $C_f$ reasonably well, but often through strong error cancellation, particularly between the turbulent torque and mean-flux contributions; individual terms can deviate by more than 20\% of $C_f$. For the hill case, where wall-resolved large-eddy simulation (WRLES) is used as the reference, errors are significantly larger. The dominant erroneous contribution differs by model and may exceed several times the local $C_f$, depending on streamwise position. In separated-flow regions, the error cancellation that was observed in the flat-plate case largely disappears for the hill case, and the leading source of error shifts between mechanisms.  
These results highlight the value of mechanism-resolved diagnostics and provide guidance for targeted turbulence-model improvements.
\end{abstract}

\section{Introduction}
Wall-bounded turbulence over complex geometries arises in many engineering and geophysical contexts, ranging from aircraft wings and turbine blades to ships, atmospheric boundary layers, and wind farms. Because high-fidelity numerical tools such as direct numerical simulation (DNS) are prohibitively costly at practically relevant Reynolds numbers \citep{choi2012grid,yang2021grid}, modeling is often the only practical option. Yet, turbulence models frequently fail to capture the intricate physics of these flows. Effects such as streamline curvature, pressure gradients, and flow separation remain challenging, and it is often unclear which aspects of the flow physics are inadequately represented in a given model \citep{huang2023distilling}. Addressing this gap requires an error-diagnostic framework that attributes modeling errors to distinct physical mechanisms arising from the underlying assumptions and calibrations.

A natural starting point for such an error-diagnostic framework is the skin-friction coefficient, $C_f$, a fundamental engineering quantity that reflects the combined influence of several underlying physical mechanisms. 
Over the past two decades, a variety of decomposition frameworks have provided insight into these processes. 
The seminal work of \cite{fukagata2002contribution} introduced the now well-known FIK decomposition, which clarified the distinct contributions of different flow processes to $C_f$. 
Other $C_f$ decompositions include formulations based on the kinetic-energy integral \citep{renard2016theoretical}, the angular-momentum integral (AMI) \citep{elnahhas2022enhancement}, extensions for rough-wall flows \citep{zhang2024integral}, and formulations for flow outside the wall layer \citep{xia2021skin,volino2018determination}. 
The AMI-based formulation, in particular, decomposes $C_f$ into contributions from viscous effects, Reynolds shear stress, total mean flux, freestream pressure gradients, and terms associated with departures from the boundary-layer approximation.
Unlike classical FIK-type relations applied to boundary layers, this approach isolates the viscous term as skin-friction of the equivalent laminar boundary layer \citep{blasius1907} in a single Reynolds-number dependent term. 
This enables a clear interpretation of the remaining terms as augmentations relative to a well-defined laminar reference state.
Furthermore, the AMI method remains well-defined and physically
interpretable even in regions of flow separation and recirculation and has been applied to flows over Gaussian bumps and airfoils with both favourable and adverse pressure gradients \citep{kianfar2025moment}. 
Despite these advances, existing formulations—including the FIK and AMI decompositions—remain largely restricted to two-dimensional mean flows, with extensions to three-dimensional (3D) configurations of engineering relevance still rare. 
Moreover, most integral decompositions have so far been used mainly for post-processing of high-fidelity data, rather than as diagnostic tools for turbulence models. 
The present study addresses these gaps by extending $C_f$ decomposition frameworks to 3D turbulent boundary layers (TBLs) and exploring their potential as an error-diagnostic tool for turbulence-model evaluation and improvement.

On the modeling side, $C_f$ has long served as a benchmark metric \citep{launder1983numerical,bose2018wall}, with emphasis placed on matching its magnitude to experimental or DNS results. In the context of wall-modeled large-eddy simulation (WMLES), various forms of log-layer mismatch can introduce 10--15\% error in predicted skin friction, motivating remedies such as random forcing \citep{deleon2019role}, filtering \citep{yang2017log,bou2004large}, displaced LES/wall-model matching \citep{KawaiLarsson2012,shin2025addressing}, among others \citep{maejima2024physics,yang2024grid,liu2025constrained}. More recently, data-driven approaches have also targeted $C_f$. For example, \cite{hu2025data} developed low-Reynolds-number corrections to the $k$--$\epsilon$ model that improve $C_f$ predictions in boundary layers with pressure gradients, \cite{bae2022scientific} applied reinforcement learning and trained wall models that capture $C_f$ in plane channel flow as well as separated periodic hill flows.
Indeed, most machine-learned closures have relied on $C_f$ as a validation quantity \citep{yang2025data,parish2016paradigm,wu2025development,ahmed2025data}. In rough-wall flows, recent reviews noted that predictive methods remain uncertain by at least 10\% \citep{Chung2021,yang2023search}, underscoring the continuing emphasis on skin-friction matching in turbulence-model development. However, while accurate $C_f$ prediction suggests improved representation of near-wall dynamics, it does not guarantee overall model fidelity: because $C_f$ reflects the combined action of multiple physical processes, the correct magnitude may arise from compensating errors among them. This study therefore seeks to bridge this gap by assessing how turbulence models represent the individual mechanisms that contribute to $C_f$.
Further, we emphasize that the present study does not challenge the established turbulence modeling framework of RANS closures, which are designed primarily to reproduce mean-flow quantities with acceptable engineering accuracy. 
Rather, we propose a mechanism-resolved diagnostic framework that complements conventional validation by revealing how individual physical processes combine to produce the predicted aggregate $C_f$. 
For instance, a model may predict the correct total $C_f$ through compensating errors, where an overprediction in one mechanism (e.g. turbulent torque) is offset by an underprediction in another (e.g. mean-flux contribution). 
Conversely, multiple small errors may compound and lead to significant deviations in skin friction. 
Identifying such distinctions is important because if a known physical deficiency contributes to error cancellation, correcting it in isolation may actually degrade the overall prediction.
In this sense, accurate $C_f$ is viewed as informative but not by itself sufficient in representing its underlying physical contributions.

The phenomenon of error cancellation has been highlighted in previous studies. For instance, \cite{klemmer2021implied} derived a transport equation for the model-form error in Reynolds-averaged Navier–Stokes (RANS) by comparing the exact Reynolds stress tensor with that modeled using the Boussinesq eddy-viscosity hypothesis. Their analysis of channel flows revealed fortuitous cancellation between the Boussinesq approximation and the modeled transport equations: an underestimation of $k$ and overestimation of $\epsilon$ combine to yield more accurate eddy viscosity $\nu_t$ in both $k$–$\epsilon$ and $k$–$\omega$ closures. Despite being fundamentally less accurate, the $k$–$\omega$ model outperformed the $k$–$\epsilon$ model in \textit{a posteriori} predictions of mean velocity profiles and turbulent shear stress, owing to more effective redistribution of errors. Extending this analysis, \cite{klemmer2023turbulence} showed similar behavior in turbulent planar jets and separation bubbles, identifying two distinct modes of error cancellation associated with wall-bounded and free-shear flows. These findings further underscore the need to quantify modeling errors in terms of distinct physical mechanisms.

In this study, we combine the AMI-based $C_f$ decomposition with high-fidelity datasets to isolate and quantify the contributions of individual flow processes to turbulence-model error. Instead of treating $C_f$ as a single scalar benchmark, we decompose it into interpretable terms that reveal which mechanisms—such as Reynolds shear stress, freestream pressure gradients, or boundary-layer growth—are misrepresented by a given model. To demonstrate this approach, we examine two flow configurations: the canonical flat-plate TBL, which serves as a baseline for establishing the methodology, and TBL flow over the asymmetric three-dimensional BeVERLI (Benchmark Validation Experiments for RANS and LES Investigations) hill \citep{gargiulo2020examination}. Previous validation studies of the symmetric BeVERLI hill have revealed significant RANS–experiment discrepancies in predicting separation and vortex shedding \citep{gargiulo2021flow,williams2022comparison,gargiulo2022computations}. Here, we extend the analysis to the asymmetric geometry, using wall-resolved large-eddy simulation (WRLES) as the high-fidelity reference and the AMI-based framework as the error-diagnostic tool. In doing so, we address two complementary challenges: the application challenge of extending $C_f$ decompositions to complex three-dimensional turbulent boundary layers, and the methodological challenge of attributing RANS model errors in skin friction to individual physical mechanisms.

The rest of the paper is organized as follows. In \S\ref{sec:2}, we present the integral analysis based on the AMI technique applied to three-dimensional flows, followed by the error-analysis framework for turbulence models and an aleatory uncertainty quantification method. 
Details of the datasets are given in \S\ref{sec:3}, including the computational domain, grid, and boundary conditions for the flat-plate turbulent boundary layer and the three-dimensional bump cases. 
Results are presented in \S\ref{sec:4}, and conclusions are drawn in \S\ref{sec:5}.


\section{Methodology} \label{sec:2}


We extend the AMI framework of \cite{elnahhas2022enhancement} to three-dimensional mean flows in \S\ref{sec:2.1} and formulate the corresponding error-analysis methodology in \S\ref{sec:2.2}. We also introduce an aleatory uncertainty quantification method in \S\ref{sec:noise_analysis}.

\subsection{AMI for three-dimensional mean flow}
\label{sec:2.1}

First, the freestream momentum equation is given by
\begin{equation}\label{eq:free}
\frac{\partial U}{\partial t}+U \frac{\partial U}{\partial x} + V \frac{\partial U}{\partial y} + W \frac{\partial U}{\partial z}=-\frac{1}{\rho} \frac{\partial P}{\partial x} + \nu\left[ \frac{\partial^2 U}{\partial x^2}+ \frac{\partial^2 U}{\partial y^2}+\frac{\partial^2 U}{\partial z^2}\right]
\end{equation}
where $U$, $V$, and $W$ denote the inviscid streamwise, spanwise, and wall-normal velocity components in the absence of a boundary layer and turbulence.
These flow variables do not satisfy the no-slip boundary condition at the wall.  
To reconstruct the streamwise freestream velocity $U$, we follow \cite{kianfar2025moment} and use the irrotational, inviscid solution from the Bernoulli equation \citep{griffin2021general},  
\begin{equation}
    U = \pm \sqrt{\frac{2}{\rho} \left( P_{o,\text{ref}} - P \right) - V^2 - W^2}, 
    \label{eq:velocity_equation}
\end{equation}
where the reference pressure $P_{o,\text{ref}}$ is taken as the maximum stagnation pressure, nearly constant in the freestream. In practice, $V$ and $W$ are taken from the corresponding viscous flow and are negligible in the freestream.  

In the presence of a boundary layer and turbulence effects, the mean continuity and streamwise momentum equations for incompressible three-dimensional flow are  
\begin{equation}\label{eq:continuity}
\frac{\partial \bar{u}}{\partial x}+\frac{\partial \bar{v}}{\partial y}+\frac{\partial \bar{w}}{\partial z}=0 
\end{equation}
\begin{equation}\label{eq:x-momentum}
\frac{\partial \bar{u}}{\partial t}+\frac{\partial \bar{u}^2}{\partial x}+\frac{\partial \bar{u} \bar{v}}{\partial y}+\frac{\partial \bar{u} \bar{w}}{\partial z} = -\frac{1}{\rho} \frac{\partial \bar{p}}{\partial x}+ \nu\left[ \frac{\partial^2 \bar{u}}{\partial x^2}+ \frac{\partial^2 \bar{u}}{\partial y^2}+\frac{\partial^2 \bar{u}}{\partial z^2}\right]-\left[\frac{\partial \overline{u^{\prime 2}}}{\partial x}+\frac{\partial \overline{u^{\prime} v^{\prime}}}{\partial y}+\frac{\partial \overline{u^{\prime} w^{\prime}}}{\partial z}\right] 
\end{equation}
where $x$, $y$ and $z$ denote streamwise, wall-normal and spanwise directions respectively, with $u$, $v$ and $w$ being the velocities in these directions.
Subtracting \eqref{eq:x-momentum} from \eqref{eq:free} and adding $(U - \bar{u})$ times \eqref{eq:continuity} gives the streamwise momentum-deficit equation:
\begin{equation}\label{eq:deficit}
\frac{\partial[\left(U-\bar{u}\right) \bar{u}]}{\partial x}
+\frac{\partial[\left(U-\bar{u}\right) \bar{v}]}{\partial y}
+\fcolorbox{red}{white}{$\mathstrut\displaystyle \frac{\partial[\left(U-\bar{u}\right)\bar{w}]}{\partial z}$}
+\left(U-\bar{u}\right) \frac{\partial U}{\partial x}
=-\nu \frac{\partial^2 \bar{u}}{\partial y^2}
+\frac{\partial \overline{u^{\prime} v^{\prime}}}{\partial y}
-I_x,
\end{equation}
where
\begin{equation}\label{eq:Ix}
\begin{split}
I_x \equiv \frac{\partial\left(U-\bar{u}\right)}{\partial t}
+ \left(V-\bar{v}\right)\frac{\partial U}{\partial y}
+ \fcolorbox{red}{white}{$\mathstrut\displaystyle\left(W-\bar{w}\right)\frac{\partial U}{\partial z}$}
+ \frac{1}{\rho} \frac{\partial\left(P-\bar{p}\right)}{\partial x}\\
- \nu\left[ \frac{\partial^2 (U-\bar{u})}{\partial x^2} +\frac{\partial^2 U}{\partial y^2}
+ \fcolorbox{red}{white}{$\mathstrut\displaystyle\frac{\partial^2 (U-\bar{u})}{\partial z^2}$} \right]
- \frac{\partial \overline{u^{\prime 2}}}{\partial x}
- \fcolorbox{red}{white}{$\mathstrut\displaystyle\frac{\partial \overline{u^{\prime}w^{\prime}}}{\partial z}$}
\end{split}
\end{equation}
denotes terms that lead to departure from boundary layer approximation.
The terms enclosed in red boxes are due to mean flow three dimensionality.
These are additional terms that extend the 2D AMI method in \cite{elnahhas2022enhancement}  to incorporate spanwise effects for application to 3D geometries.
It is also important to note that both equations \eqref{eq:free} and \eqref{eq:x-momentum} are evaluated at the same spatial locations, the former reconstructed from the inviscid solution and the latter valid throughout the domain, so the subtraction is performed locally within the same physical domain rather than between disjoint regions.

The AMI equation can then be constructed by multiplying \eqref{eq:deficit} by $(y-\ell)/(U_{io}^2 \ell)$ and integrating from $0$ to $\infty$:
\begin{align} \label{eq:AMI}
    \frac{C_f}{2} &= \frac{1}{Re_{\ell}} 
    + \int_0^{\infty} \frac{-\overline{u^{\prime} v^{\prime}}}{U_{io}^2 \ell} \, \mathrm{d}y 
    + \frac{\partial \theta_{\ell x}}{\partial x} 
    - \frac{\theta_x - \theta_{\ell_x}}{\ell} \frac{\mathrm{d}\ell}{\mathrm{d}x} \nonumber 
     + \fcolorbox{red}{white}{$\mathstrut\displaystyle\frac{\partial \theta_{\ell z}}{\partial z}$}
     - \fcolorbox{red}{white}{$\mathstrut\displaystyle\frac{\theta_z - \theta_{\ell_z}}{\ell}\frac{\mathrm{d}\ell}{\mathrm{d}z}$} 
        \\ & \quad
    + \frac{\theta_v}{\ell} 
    + \frac{\delta_{\ell_x}^* + 2 \theta_{\ell_x}}{U_{io}} \frac{\partial U}{\partial x} 
    + \fcolorbox{red}{white}{$\mathstrut\displaystyle\frac{2 \theta_{\ell_z}}{U_{io}}\frac{\partial U}{\partial z}$}  
    + \mathcal{I}^{\ell}.
\end{align}
where
\begin{equation}
\quad R e_{\ell}=U_{io} \ell / v,~~
\mathcal{I}^{\ell} \equiv \int_0^{\infty}\left(1-\frac{y}{\ell}\right) \frac{I_x}{U_{io}^2} \mathrm{~d} y
\end{equation}
and the normalizing velocity $U_{io}$ comes from the irrotational solution when evaluated at the wall $U_{io} = U(y=0)$  \citep{kianfar2023quantifying}. 
The boundary layer momentum thicknesses are: 
\begin{equation}
 \quad \theta_x \equiv \int_0^{\infty}\left(\frac{U-\bar{u}(y)}{U_{io}}\right) \frac{\bar{u}(y)}{U_{io}} \mathrm{d} y, \quad \theta_z \equiv \int_0^{\infty}\left(\frac{U-\bar{u}(y)}{U_{io}}\right) \frac{\bar{w}(y)}{U_{io}} \mathrm{d} y
\end{equation}

The momentum and displacement thicknesses based on the length scale $\ell$ are:

\begin{equation}
\begin{aligned}
    \delta_{\ell_x}^* &\equiv \int_{0}^{\infty} \left(1 - \frac{y}{\ell}\right) \left(\frac{U - \bar{u}}{U_{io}}\right) \, \mathrm{d}y, \\
    \theta_{\ell_x} &\equiv \int_{0}^{\infty} \left(1 - \frac{y}{\ell}\right) \frac{\bar{u}}{U_{io}} \left(\frac{U - \bar{u}}{U_{io}}\right) \, \mathrm{d}y,\\
    \theta_{\ell_z} &\equiv \int_{0}^{\infty} \left(1 - \frac{y}{\ell}\right) \frac{\bar{w}}{U_{io}} \left(\frac{U - \bar{u}}{U_{io}}\right) \, \mathrm{d}y,\\
    \theta_{v} &\equiv \int_{0}^{\infty} \left(1 - \frac{y}{\ell}\right) \frac{\bar{v}}{U_{io}} \left(\frac{U - \bar{u}}{U_{io}}\right) \, \mathrm{d}y,\\
\end{aligned}
\label{eq:label}
\end{equation}
The length scale $\ell$ can be thought of as the point about which these torques due to various $C_f$ contributions act on the TBL.
Its value is chosen to be 4.54 times the streamwise momentum thickness ($\theta_x$) following \cite{kianfar2025moment}.

Equation \eqref{eq:AMI} constitutes the AMI decomposition of the skin-friction coefficient,  
\begin{equation}\label{eq:cf}
C_f \equiv \frac{\tau_w}{\tfrac{1}{2} \rho U_{io}^2},
\end{equation}
where $\tau_w$ is the wall shear stress. 
The first term on the right-hand side of \eqref{eq:AMI} represents the viscous contribution, obtained by comparing the turbulent boundary layer with an equivalent laminar boundary layer at the same Reynolds number. 
This $1/Re_{\ell}$ term is also referred to as the `laminar friction' term \citep{elnahhas2022enhancement, kianfar2025moment}, corresponding to the wall shear that would arise from molecular diffusion alone in the absence of turbulent momentum transport. The length scale 
$\ell$ denotes the center of action of this viscous contribution in the reference laminar BL and does not imply the existence of a laminar region within the TBL.
The second term quantifies the turbulent torque, capturing the enhancement of wall shear stress due to momentum transfer by turbulent fluctuations. 
The following five terms arise from mean-flow fluxes associated with streamwise and spanwise growth of the boundary layer; these typically reduce skin friction, as boundary-layer thickening carries momentum away from the wall and offsets turbulent enhancement. 
The next two terms represent freestream pressure-gradient effects: a favourable pressure gradient increases $C_f$ by accelerating the freestream, whereas an adverse gradient reduces it. 
Finally, residual contributions due to unsteadiness, three-dimensionality, or other departures from ideal two-dimensional conditions are collected into $\mathcal{I}^{\ell}$, which is negligible for steady, zero-pressure-gradient flat-plate flows.

Alternative but mathematically equivalent regroupings of the terms in \eqref{eq:AMI} are possible. 
These are discussed in appendix \ref{appA}, where we demonstrate that the principal qualitative trends remain robust under such rearrangements.

\subsection{AMI-enabled error analysis}\label{sec:2.2}

The AMI framework enables a term-by-term decomposition of modeling error by associating each contribution to $C_f$ with a distinct physical mechanism. More generally, any quantity of interest (QoI) may be expressed as the sum of such contributions. For a high-fidelity reference solution, such as DNS or WRLES, we may write
\begin{equation}
    \mathrm{QoI}_{\mathrm{ref}} = \sum_{i} C_i^{\mathrm{ref}},
\end{equation}
where $C_i^{\mathrm{ref}}$ denotes the contribution of the $i$th physical mechanism, evaluated from the reference flow field. For a lower-fidelity model prediction, e.g.,  RANS, the same QoI can be expressed as
\begin{equation}
    \mathrm{QoI}_{\mathrm{M}} = \sum_{i} C_i^{\mathrm{M}},
\end{equation}
where $C_i^{\mathrm{M}}$ is the corresponding contribution obtained from the modeled flow field.
The discrepancy between the two is therefore
\begin{equation}
    \mathrm{QoI}_{\mathrm{M}} - \mathrm{QoI}_{\mathrm{ref}} = \sum_{i} \left( C_i^{\mathrm{M}} - C_i^{\mathrm{ref}} \right),
\end{equation}
showing explicitly that the total error arises from the sum of term-by-term differences in the underlying physical contributions. This additive structure naturally motivates an error-budget interpretation: each mechanism contributes a separate component of error, which may reinforce or cancel others.  

Here, the QoI is the skin-friction coefficient $C_f$, and its decomposition is given by the AMI:
\begin{align}\label{eq:error_analysis}
\frac{\Delta C_f}{2} ={}&
\underbrace{\Delta \left( \frac{1}{Re_\ell} \right)}_{\text{Laminar friction}}
+ \underbrace{\Delta \left( \int_0^{\infty} \frac{-\overline{u^{\prime} v^{\prime}}}{U_{io}^2 \ell} \, \mathrm{d}y \right)}_{\text{Turbulent torque}} \nonumber \\
&+ \underbrace{\Delta\left( \frac{\partial \theta_{\ell x}}{\partial x}
    - \frac{\theta_x - \theta_{\ell_x}}{\ell} \frac{\mathrm{d}\ell}{\mathrm{d}x}
    + \frac{\partial \theta_{\ell z}}{\partial z}
    - \frac{\theta_z - \theta_{\ell_z}}{\ell} \frac{\mathrm{d}\ell}{\mathrm{d}z}
    + \frac{\theta_v}{\ell} \right)}_{\text{Total mean flux}} \nonumber \\
&+ \underbrace{\Delta\left( \frac{\delta_{\ell_x}^* + 2 \theta_{\ell_x}}{U_{io}} \frac{\partial U}{\partial x}
+ \frac{2 \theta_{\ell_z}}{U_{io}} \frac{\partial U}{\partial z} \right)}_{\text{Pressure-gradient effects}}
+ \underbrace{\Delta\left( \mathcal{I}^{\ell} \right)}_{\text{Departure from BL approximation}}.
\end{align}
Here, $\Delta$ denotes the difference between the turbulence-model prediction and the high-fidelity reference. For example, the error in $C_f$ from the $k-\omega$ SST model relative to WRLES is  
\[
\Delta C_{f,\text{SST}}/2 = C_{f,\text{SST}}/2 - C_{f,\text{WRLES}}/2.
\]
Equation \eqref{eq:error_analysis} thus provides a systematic decomposition of skin-friction error into contributions from individual physical processes. In this study, we apply this framework to five RANS models, i.e., the two-equation $k-\omega$ SST model \citep{Menter_1994}, Chien's $k$--$\epsilon$ model \citep{Chien_1982}, the one-equation Spalart--Allmaras model \citep{spalart1988direct}, the four-equation $v^2-f$\citep{laurence2005robust} and the Reynolds-stress-based SSG--LRR model \citep{Eisfeld_2016}, and two flow configurations, i.e., the flat-plate turbulent boundary layer and the BeVERLI hill.

\subsection{Aleatory uncertainty quantification} \label{sec:noise_analysis}

Previous studies \citep{kianfar2023quantifying,kianfar2025moment} emphasized that incomplete statistical convergence of turbulent boundary-layer data introduces errors that directly affect AMI analyses. For the zero-pressure-gradient flat-plate DNS of \cite{wu2017transitional}, the AMI equation balanced the skin-friction coefficient with an average discrepancy of 1.4\%. This deviation was traced to the amplification of statistical noise by streamwise derivatives, which are especially sensitive to finite sampling \citep{kianfar2023quantifying}. In \cite{kianfar2025moment}, a normalized residual was introduced and shown to increase in regions of strong pressure gradients, growing markedly near incipient separation where unsteady dynamics produce noisier statistics. Both studies identified streamwise derivatives as the dominant source of error amplification, with additional contributions from wall-normal integrals and edge definitions.  

To systematically assess the susceptibility of AMI terms to convergence errors, we introduce controlled synthetic perturbations that emulate noise in the data. The perturbations are constructed to be divergence-free, to have a prescribed magnitude and to obey a Kolmogorov $k^{-5/3}$ spectrum. The resulting noisy velocity fields are then used to recompute the AMI terms, and the differences provide estimates of the sensitivity of each contribution to sampling error. In particular, adding noise to converged RANS solutions offers a clean baseline, since RANS solutions are free from temporal-averaging error.  

The procedure is as follows. Consider two-dimensional mean flow. Incompressibility is enforced by introducing a scalar stream function $\psi(x,y)$, with perturbation velocities
\begin{equation}
e'_u=\frac{\partial \psi}{\partial y},\qquad
e'_v=-\frac{\partial \psi}{\partial x}.
\label{eq:uv_from_psi}
\end{equation}
In Fourier space on a periodic domain $(L_x,L_y)$ with wavenumbers $k_x=2\pi n_x/L_x$, $k_y=2\pi n_y/L_y$ and $\boldsymbol{k}=(k_x,k_y)$, this becomes
\begin{equation}
\widehat{e'_u}(\boldsymbol{k}) = i\,k_y\,\widehat{\psi}(\boldsymbol{k}),\qquad
\widehat{e'_v}(\boldsymbol{k}) = -i\,k_x\,\widehat{\psi}(\boldsymbol{k}),
\label{eq:uv_in_fourier}
\end{equation}
so that $\boldsymbol{k}\cdot\widehat{\boldsymbol{e'}}(\boldsymbol{k})=0$, ensuring the divergence-free constraint.  
To enforce the $E(\kappa)\propto \kappa^{-5/3}$ spectrum, where $\kappa=\sqrt{k_x^2+k_y^2}$, we prescribe a random-phase stream function field with Fourier amplitudes
\begin{equation}
\widehat{\psi}(\boldsymbol{k}) \;=\; A(\kappa)\,e^{i\phi(\boldsymbol{k})},\qquad
A(\kappa)\propto \kappa^{-11/6},
\label{eq:psi_amp}
\end{equation}
where $\phi(\boldsymbol{k})\in [0,2\pi)$ are independent phases and $\widehat{\psi}(\mathbf{0})=0$. Because $|\widehat{e'_u}|^2+|\widehat{e'_v}|^2 \sim \kappa^2\,|\widehat{\psi}|^2$, this construction yields the desired $\kappa^{-5/3}$ velocity spectrum.  
An inverse Fast Fourier transform (FFT) of \eqref{eq:uv_in_fourier} produces the perturbations, which are interpolated onto the computational grid. Their root-mean-square is defined as
\begin{equation}
e'_{\mathrm{rms}} \;=\; \left\langle\, (e'_u)^2+(e'_v)^2 \,\right\rangle^{1/2},
\end{equation}
and the perturbations are rescaled as $(e'_u,e'_v)\rightarrow \alpha(e'_u,e'_v)$, where $\alpha$ is chosen so that the prescribed noise level corresponds to a specified fraction of the reference velocity $U_\infty$. The final noisy field is then
\begin{equation}
\widetilde{\boldsymbol{u}}(x,y) \;=\; \overline{\boldsymbol{u}}(x,y) \;+\; \boldsymbol{e}'(x,y),
\end{equation}
where $\overline{\boldsymbol{u}}$ is the baseline mean solution and $\boldsymbol{e}'$ is the divergence-free perturbation defined above.

\section{Datasets and Computational Details} \label{sec:3}

We will apply our diagnostics to the flat-plate turbulent boundary layer and BeVERLI hill.
High fidelity DNS data of the flat-plate turbulent boundary layer is readily available in \cite{towne2023database}.
We will generate high-fidelity WRLES data for BeVERLI hill.
In addition, we will generate low-fidelity RANS data for both flat-plate turbulent boundary layer and BeVERLI hill for the one-equation SA model, the two-equation $k-\omega$ SST and Chien's $k-\epsilon$ models,  the four-equation $v^2-f$ model and the seven-equation SSG-LRR FRSM model. 
This section details the simulation setup.

\subsection{Flat-plate boundary layer}
\subsubsection{Reference DNS}
We utilize reference data from the incompressible zero-pressure-gradient flat-plate turbulent boundary layer direct numerical simulations in \cite{towne2023database}.
More specifically, the BL1 dataset is utilized and covers friction Reynolds number from $Re_\tau \approx 292 - 729$.
The computational domain spans $L_x = 450 \theta_{avg}$, $L_y = 50 \theta_{avg}$ and $L_z = 70 \theta_{avg}$ in the streamwise, wall-normal and spanwise directions. 
Here $\theta_{avg}$ denotes the momentum thickness averaged along the streamwise direction. 
Further details regarding grid-resolution and boundary conditions can be found in \cite{schlatter2010assessment} and \cite{towne2023database} respectively.
The resulting mean flow-field was averaged through 26 eddy turnover times (after transients) and is used to compute the reference AMI decompositions in the current study.

\subsubsection{RANS}

For the flat plate boundary layer, a 2D Kelbnoff flat plate configuration \citep{klebanoff1954characteristics} is considered. 
A uniform velocity is prescribed at the domain inlet, with a slip boundary between the inlet and the beginning of the wall following the setup in \cite{rumsey2018tmr}. 
The outlet is modeled as a fixed pressure outlet, and the remaining boundary conditions are modeled as slip boundary conditions. 
The grid spacing in the wall normal direction is progressively increased with a hyperbolic tangent distribution while the first cell centroid is at $y^+ <1$. 
A finer grid spacing is opted near the starting edge of the plate to capture large gradients due to the change from no slip to slip boundary condition again following \cite{rumsey2018tmr}. 

For the RANS simulations, the NPHASE-PSU code \citep{Kunz_2001} is primarily used for all models except the $v^2-f$ model.
The in-house CFD solver employs segregated pressure-based finite-volume methods with its numerics validated in several prior studies \citep{jain2022study,jain2022second,jain2023assessment}. 
The momentum equations are discretized using a second-order upwind scheme, while turbulence scalars are discretized with a hybrid scheme. The second-moment closure equations and the $\omega$-equation in the SSG-LRR model use a first-order upwind scheme for stability.
The turbulence models used include the Menter $k-\omega$ SST model \citep{Menter_1994}, the Chien $k$–$\epsilon$ model \citep{Chien_1982}, and the $\omega$-based SSG-LRR full Reynolds stress model (FRSM) \citep{Eisfeld_2016}. 
The $v^2-f$ model \citep{laurence2005robust} was computed using the SIMPLE method in OpenFOAM \citep{jasak2009openfoam}.
All models follow implementations as detailed on the NASA Turbulence Modeling Resource site \citep{rumsey2010description}.
These implementations use the standard fully-turbulent formulations \citep{rumsey2010description, jespersen2016overflow}  for the zero pressure gradient TBL and do not model the laminar to turbulent transition. 

\subsection{BeVERLI hill}

\subsubsection{WRLES}

\begin{figure}
\centerline{\includegraphics[width=0.4\linewidth]{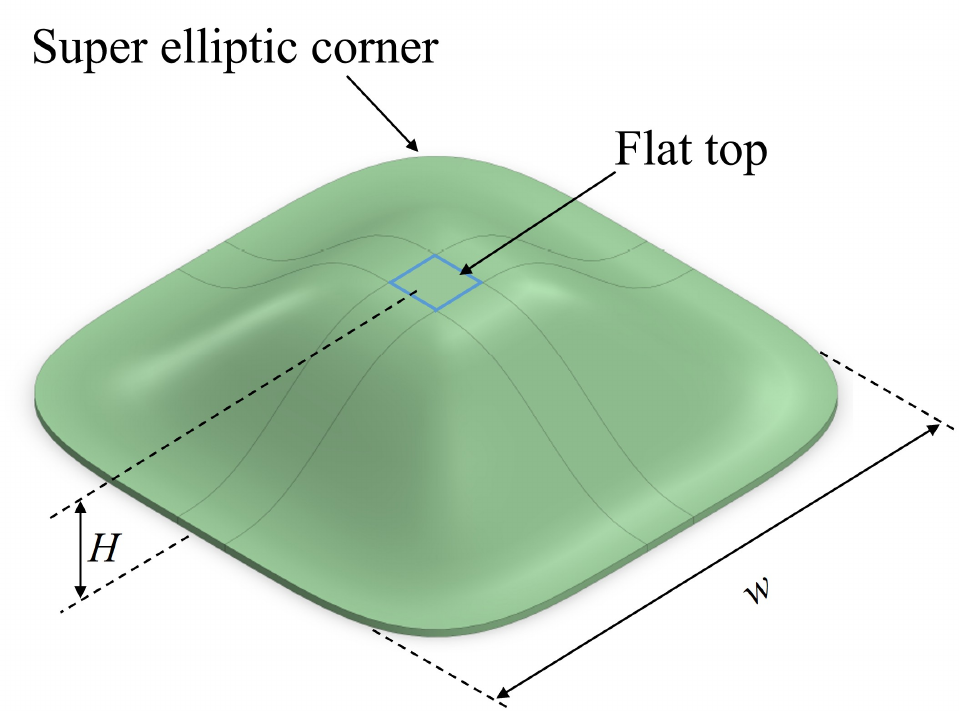}}
  \caption{BeVERLI hill geometry. Here, $H$ and $w$ denote the hill height and width respectively, with aspect ratio $w/H = 5$. Further details can be found in \cite{gargiulo2020examination,gargiulo2023strategies}.}
\label{fig:geometry}
\end{figure}

\begin{figure}
\centerline{\includegraphics[width=\linewidth]{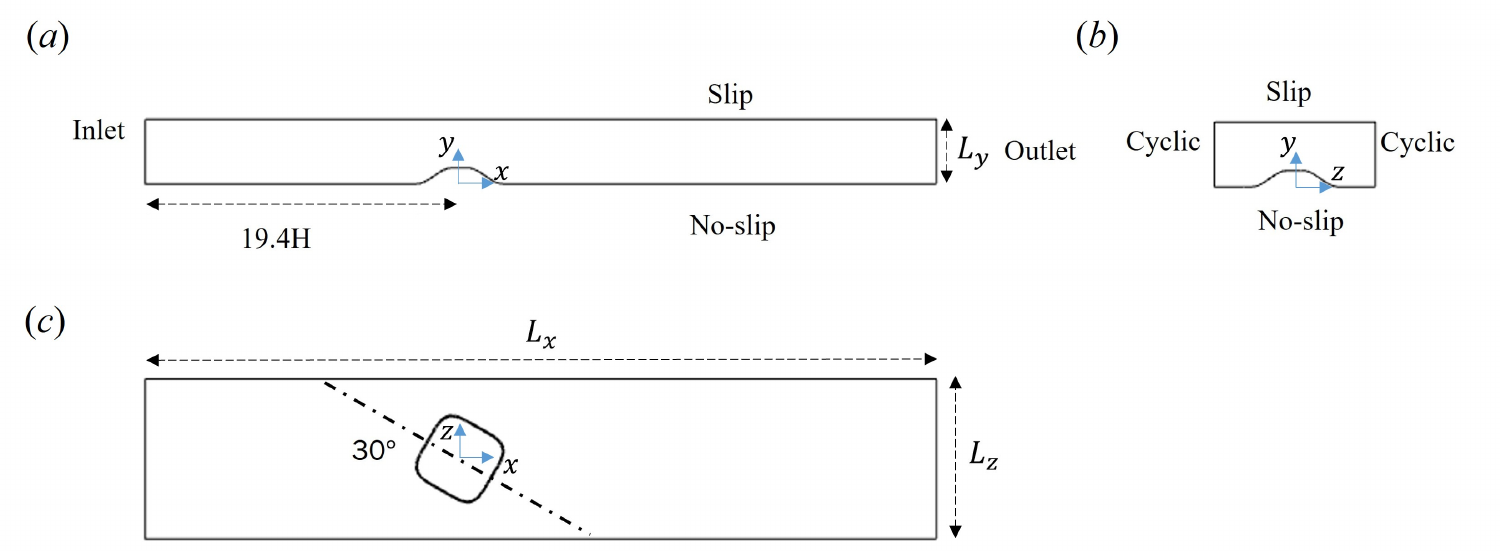}}
  \caption{Computational domain and boundary conditions shown in: (a) spanwise plane ($z/H=0$), (b) streamwise plane ($x/H=0$), and (c) wall-normal plane ($y/H=0$).}
\label{fig:bc}
\end{figure}

The configuration considered is the BeVERLI hill, a three-dimensional bump that has served as a benchmark for RANS validation and experimental studies \citep{gargiulo2020examination,gargiulo2021flow,gargiulo2022computations}. 
The hill geometry, shown in figure \ref{fig:geometry}, is defined by mirrored fifth-degree polynomials and exhibits $90^\circ$ rotational symmetry. 
Following \cite{gargiulo2020examination}, the hill is oriented at $30^\circ$ to the streamwise direction. 
The computational domain extends $19.4H$ upstream and $29.4H$ downstream of the hill, with a spanwise width of $9.9H$.
The vertical domain height $L_y$ extends to $4H$.
The hill is centered in the domain, as illustrated in figure \ref{fig:bc}. The inlet length ensures sufficient development of inflow turbulence and recovery of the log law at least $6H$ upstream of the hill.

The incompressible, unsteady Navier–Stokes equations are solved using the PIMPLE algorithm in OpenFOAM. A Wall-Adapting Local Eddy-viscosity (WALE) subgrid-scale model \citep{nicoud1999subgrid} accounts for unresolved turbulence. Numerical discretization employs second-order backward differencing for time advancement and a Gauss linear scheme for spatial gradient and divergence operators. The simulation covers seven flow-through times ($L_x/U_{\text{ref}}$), with the last five used for collecting statistics. This corresponds to approximately $240\,H/U_{\text{ref}}$ time units, consistent with prior LES studies \citep{garcia2009large,zhang2024integral}.

\begin{figure}
\centerline{\includegraphics[width=0.45\linewidth]{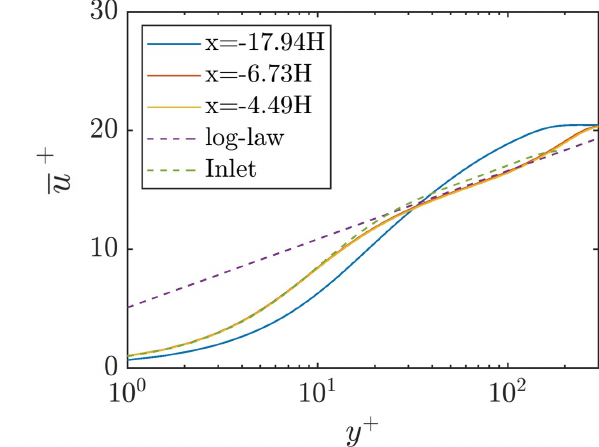}}
  \caption{Mean velocity profiles at the inlet and several $x$ locations upstream of the hill. The log-law reference is $U^+ = (1/0.4)\,\log(y^+)+5.1$.}
\label{fig:inlet_pro}
\end{figure}

Boundary conditions include a no-slip wall at the bottom surface, slip at the top, and cyclic conditions at the lateral boundaries. The outlet employs an advective velocity condition with fixed pressure, while the inlet and top use zero-gradient pressure. Inflow turbulence is generated using the Divergence-Free Synthetic Eddy Method (DFSEM) \citep{poletto2013new}, driven by velocity and Reynolds stress profiles from the DNS of \cite{lee2015direct} at friction Reynolds number $Re_{\tau}=182$. The inflow profile and its streamwise development are shown in figure \ref{fig:inlet_pro}, which illustrates that at least $6H$ upstream distance is needed to recover the log law before the hill.

\begin{figure}
\centerline{\includegraphics[width=\linewidth]{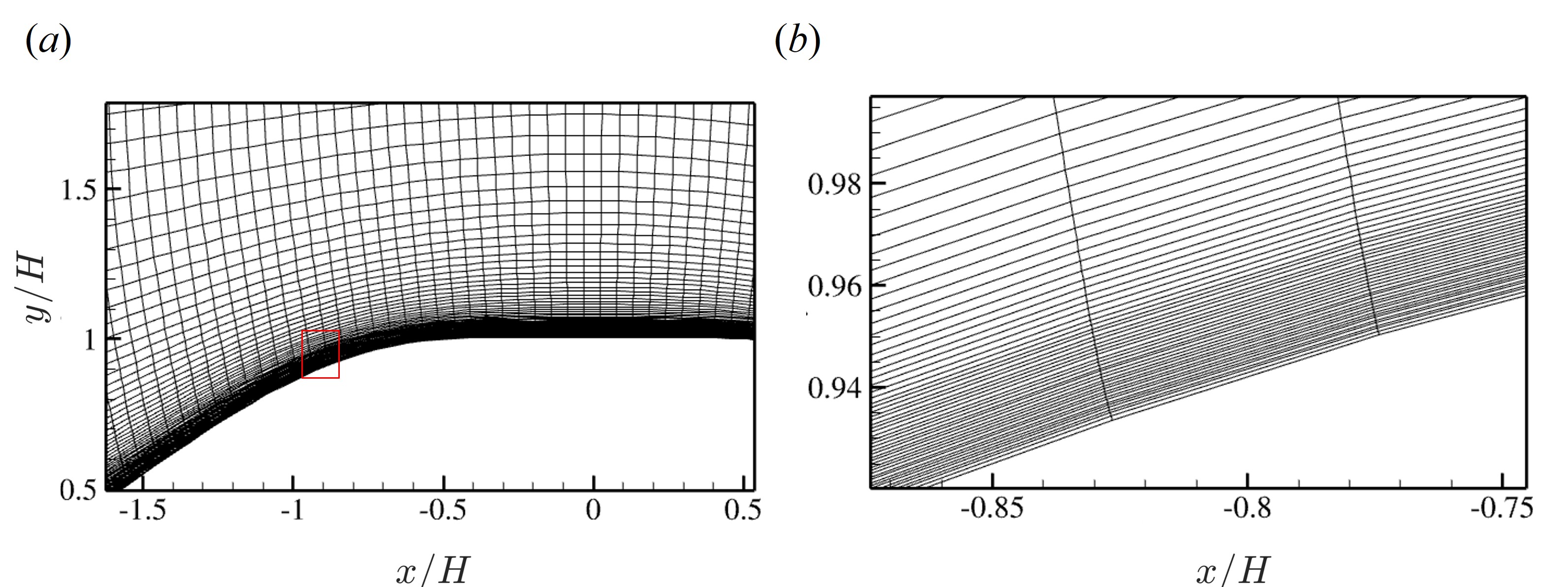}}
  \caption{Computational grid near the hill front: (a) distribution in the spanwise plane $z/H=0$; (b) close-up view of the marked region.}
\label{fig:grid}
\end{figure}

The structured grid is orthogonal to the hill surface, as shown in figure \ref{fig:grid}. At the inlet, spacings are $\Delta x^+ = 18$, $\Delta z^+ = 12$, and $\Delta y^+ < 1$ at the first off-wall cell, with stretching applied in the streamwise and wall-normal directions. The grid comprises $N_x \times N_y \times N_z = 940 \times 95 \times 662$ cells, or 58.3 million points in total.  

\begin{figure}
\centerline{\includegraphics[width=\linewidth]{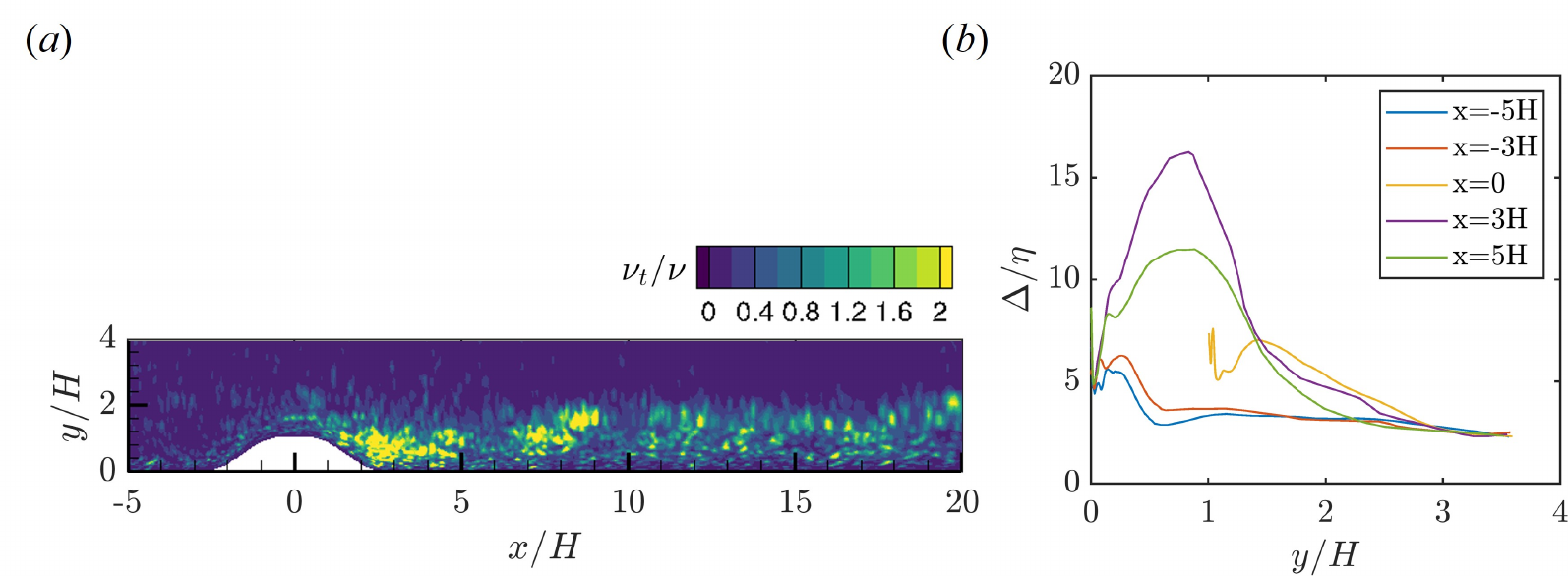}}
  \caption{Grid resolution assessment from the spanwise mid-plane $z/H = 0$: \\ (a) Instantaneous turbulent-to-molecular viscosity ratio ($\nu_t/\nu$), and (b) representative grid spacing to Kolmogorov length scale ratio ($\Delta/\eta$) at various streamwise locations.}
\label{fig:eddy_visc}
\end{figure}

Grid sufficiency is assessed using the turbulent-to-molecular viscosity ratio $\nu_t/\nu$, shown in figure \ref{fig:eddy_visc}(a). Here, this ratio is computed from instantaneous flow fields rather than mean fields, since averaging tends to smooth intermittent events and may underestimate resolution demands \citep{yang2021grid,chen2023quantifying}: intermittent structures such as shear-layer roll-up, near-wall bursting, and tip vortices impose the strongest requirements on grid resolution, and their effects are only visible in instantaneous fields. 
The low values of $\nu_t/\nu$ observed in figure \ref{fig:eddy_visc}(a) indicate that the grid is fine enough to resolve most near-wall turbulent structures, approaching DNS resolution levels. For context, DNS of separation bubbles reports $\nu_t/\nu \approx 150$–$250$ \citep{coleman2018numerical}.
While the ratio $\nu_t/\nu$ provides a useful qualitative indicator, a more stringent and physically grounded grid resolution metric is the ratio of grid spacing to the Kolmogorov length scale ($\eta$). 
Figure ~\ref{fig:eddy_visc}(b) therefore presents the distribution of $\Delta/\eta$, where the representative grid spacing is defined as $\Delta = (\Delta x \Delta y \Delta z)^{1/3}$. 
The ratio $\Delta/\eta$ remains $\mathcal{O}(3-5)$ over most of the domain, with localized peaks up within the separated shear layer, which is typical of wall-resolved LES of separated flows.
For comparison, coarse grid DNS of the wall-mounted hump configuration report $\Delta x/\eta$ and $\Delta z/\eta$ of $\mathcal{O}(10)$ in \cite{postl2006direct}.


\subsubsection{RANS}

For the RANS simulations of the BeVERLI hill, the computational domain shown in figure \ref{fig:bc} was truncated at $x=-6.73H$ upstream of the hill. This location was sufficiently far from the hill for the boundary-layer profile from the WRLES to be fully developed. The velocity profile at this plane was extracted from the WRLES and prescribed as the RANS inlet condition. At the outlet, a fixed-pressure boundary condition was applied. The bottom wall was treated with a no-slip condition, with wall shear stress computed using second-order discretization, while all remaining boundaries were modeled as slip walls. The grid resolution matched that of the WRLES configuration described above, and the same solver setup was used to run the five RANS closures: Spalart–Allmaras, $k$–$\epsilon$, $k-\omega$ SST, $v^2-f$ and the SSG–LRR Reynolds stress model.

\section{Results} \label{sec:4}

We apply the error diagnostics to the zero-pressure-gradient flat-plate turbulent boundary layer and to the BeVERLI hill, with results presented in \S\ref{sec:4.1} and \S\ref{sec:4.3} respectively. Between these, \S\ref{sec:4.2} presents a conventional validation study, highlighting the insights obtainable without the proposed error diagnostics.

\subsection{Model error diagnostics for flat-plate TBL}\label{sec:4.1}

\begin{figure}
\centerline{\includegraphics[width=\linewidth]{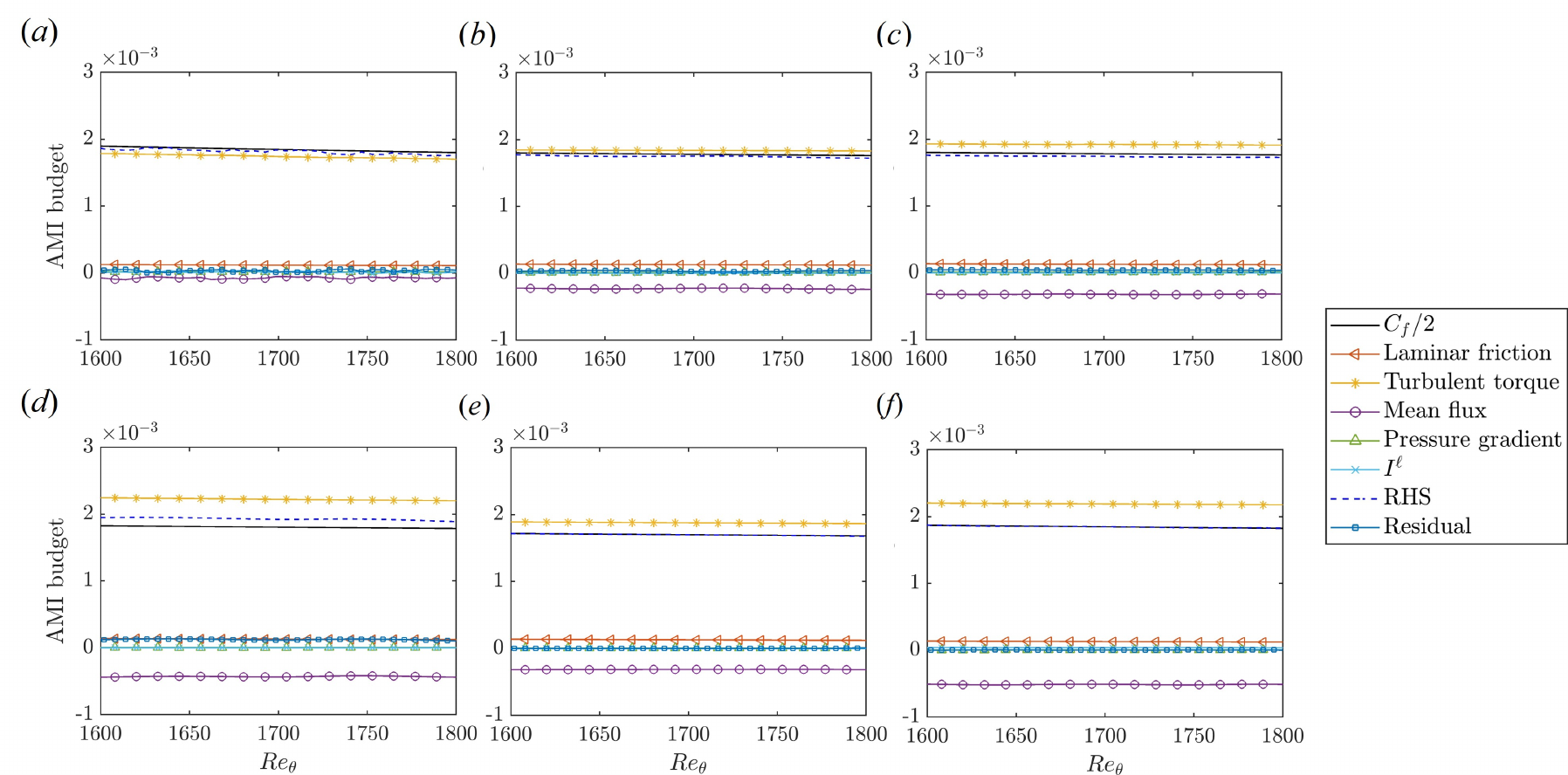}}
  \caption{AMI decomposition of $C_f$ according to \eqref{eq:AMI} for a flat-plate TBL: (a) DNS, (b) $k-\omega$ SST, (c) SA, (d) $k-\epsilon$, (e) $v^2$--$f$, and (f) SSG–LRR FRSM. Here, $C_f/2$ is evaluated directly at the wall, RHS denotes the sum of the terms on the right hand side of \eqref{eq:AMI}.}
\label{fig:bl_ami}
\end{figure}

We begin with the AMI decomposition of the skin-friction coefficient $C_f$ for the flat-plate turbulent boundary layer. Figure~\ref{fig:bl_ami} shows the decomposition from \eqref{eq:AMI}, applied to DNS data \cite{towne2023database} and to RANS results from five turbulence models, evaluated at locations sufficiently far from the inlet and outlet to avoid boundary effects. 
While the DNS employs a recycling--rescaling inflow \citep{lund1998generation}, the RANS simulations use a standard upstream inlet \citep{rumsey2018tmr} and include a finite development length before the flat-plate leading edge, allowing the turbulent boundary layer to form from the prescribed freestream conditions. 
In the present analysis, we focus on streamwise locations sufficiently far downstream from the leading edge where sensitivity to inflow conditions is
reduced, though some residual differences may reflect this distinction.

The analysis spans a range of momentum-thickness Reynolds numbers, $Re_\theta$. We observe the following. 
First, the $C_f/2$ predicted by the RANS models agrees reasonably well with DNS with a slight underprediction. 
The $k$--$\epsilon$ model shows a mild overprediction in the RHS sum of \eqref{eq:AMI} whereas the remaining models show near-consistent agreement between the RHS and their respective $C_f/2$ values.
Second, the contributions of the individual terms to $C_f$ vary only weakly with $Re_\theta$, and the models redistribute them differently relative to DNS and to one another.  
Third, in all cases, turbulent torque remains the dominant contributor to $C_f$, consistent with earlier studies \citep{elnahhas2022enhancement,de2016skin}. 
In the SA, $k$--$\epsilon$, $v^2-f$ and SSG-LRR FRSM models, this contribution overshoots $C_f/2$, a behavior also reported in AMI analyses of other TBL datasets \citep{elnahhas2022enhancement,sillero2013one}.
By contrast, the mean-flux term reduces skin friction, with the SSG-LRR FRSM model showing the strongest suppression, about $-25\%$ of $C_f/2$.  
Lastly, the viscous contribution is small (less than $5\%$ of $C_f/2$), while pressure-gradient and boundary-layer-departure effects are negligible, as expected for zero-pressure-gradient flow.

\begin{figure}
\centerline{\includegraphics[width=0.45\linewidth]{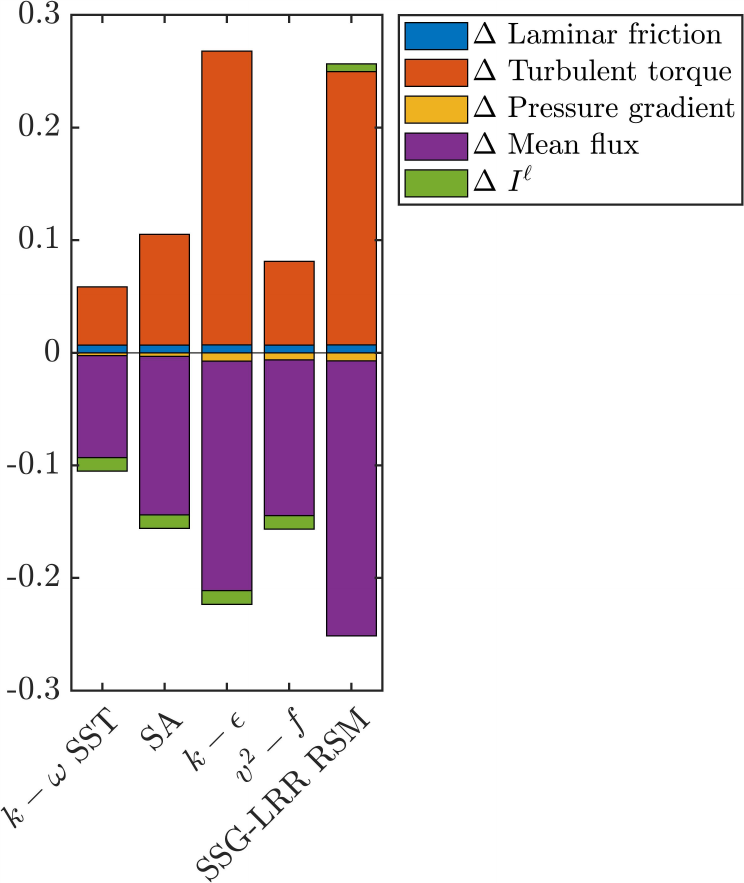}}
  \caption{Error analysis for flat-plate turbulent boundary layer at $Re_\theta=1700$, relative to DNS, for different turbulence models. Each error is normalized by $C_{f,\text{DNS}}/2$.}
\label{fig:bl_Cf_decomp_percent}
\end{figure}

While figure~\ref{fig:bl_ami} illustrates the composition of skin friction across physical mechanisms, it does not reveal how accurately each mechanism is captured in RANS. To assess this, we employ the error decomposition in \eqref{eq:error_analysis}. Figure~\ref{fig:bl_Cf_decomp_percent} shows the decomposition errors at $Re_\theta=1700$, normalized by $C_{f,\text{DNS}}/2$. The turbulent torque contribution is consistently overpredicted by all models, with its error largely cancelled by the total mean-flux contribution, a term that reflects streamwise and spanwise growth of angular momentum redistributed by mean wall-normal transport. Errors in viscous effects, pressure-gradient effects, and departures from the boundary-layer approximation are negligible. 
Importantly, individual contributions can deviate by as much as 25\% of $C_f/2$, yet these errors compensate to yield net $C_f/2$ differences of only about 5\%. This indicates that models may achieve correct $C_f$ for the wrong reasons, through misallocation of the physical budget of skin friction. As will be shown in the next subsection, while the errors in mean flux and turbulent torque cancel for the flat plate, they compound in more complex flow scenarios.

\begin{figure}
\centerline{\includegraphics[width=0.78\linewidth]{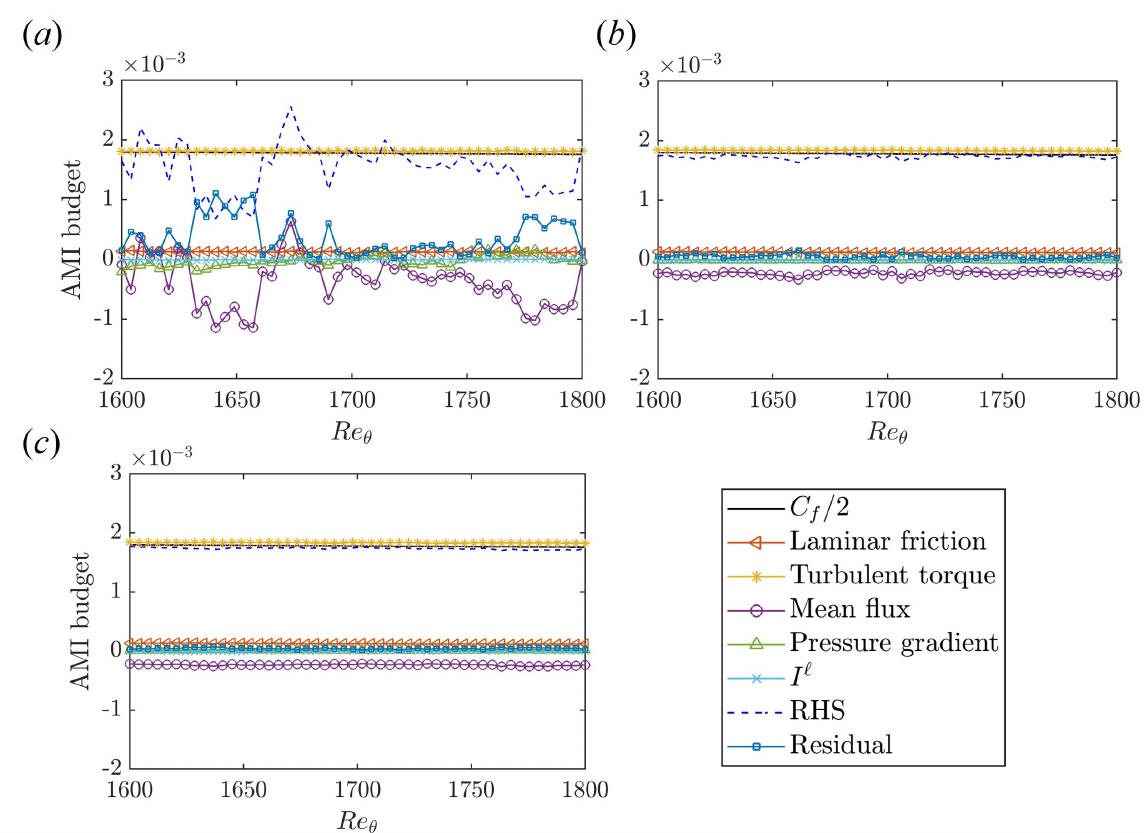}}
  \caption{Uncertainty analysis for the flat-plate $k-\omega$ SST solution with synthetic perturbations: (a) $u_{\text{noise}}/U_{\text{ref}} = 0.001$, (b) $u_{\text{noise}}/U_{\text{ref}} = 0.0001$, (c) $u_{\text{noise}}/U_{\text{ref}} = 0.00001$. Here $u_{\text{noise}}$ denotes the root-mean-square of the perturbation and $U_{\text{ref}}$ is the freestream velocity.}
\label{fig:bl_sst_noise}
\end{figure}

The DNS data for the flat plate are sufficiently converged to yield accurate integrals, allowing us to test the sensitivity of the AMI framework to statistical convergence errors. As outlined in \S\ref{sec:noise_analysis}, we introduce synthetic, divergence-free noise with a $-5/3$ Kolmogorov spectrum into the mean fields to mimic sampling errors from finite averaging. Figure~\ref{fig:bl_sst_noise} shows the resulting perturbation analysis applied to the $k-\omega$ SST model; results for the other models are qualitatively similar and are omitted for brevity. Perturbations of 0.1\%, 0.01\%, and 0.001\% of $U_{\text{ref}}$ were added to the mean fields. As shown in figure~\ref{fig:bl_sst_noise}(a), even a 0.1\% perturbation produces substantial changes in the AMI integrals. Smaller perturbations of 0.01\% and 0.001\% lead to correspondingly weaker but still noticeable effects, demonstrating that the AMI analysis is highly sensitive to sampling noise. Comparable fluctuations in mean-flux and pressure-gradient terms were also reported in AMI studies of Gaussian-bump boundary layers \citep{balin2021direct,kianfar2025moment}. For context, uncertainties in DNS datasets are typically $O(0.1\%)$ in mean velocity and wall shear stress, rising to $\sim 7\%$ for higher-order statistics such as skewness, kurtosis, and dissipation due to wall-normal resolution and discretization errors \citep{oliver2014estimating,chen2023quantifying}. 
These comparisons provide the necessary background for the analysis of the BeVERLI hill results.

\subsection{Conventional validation study for BeVERLI hill}\label{sec:4.2}

To ground the subsequent error analysis \S \ref{sec:4.3} results, we present conventional model validation analysis for the BeVERLI hill case.
We first summarize the flow phenomenology by reporting the WRLES results for the BeVERLI hill configuration.
Next, we present RANS results and compare the RANS results to the high-fidelity WRLES results.

\subsubsection{Flow phenomenology}

\begin{figure}
\centerline{\includegraphics[width=0.8\linewidth]{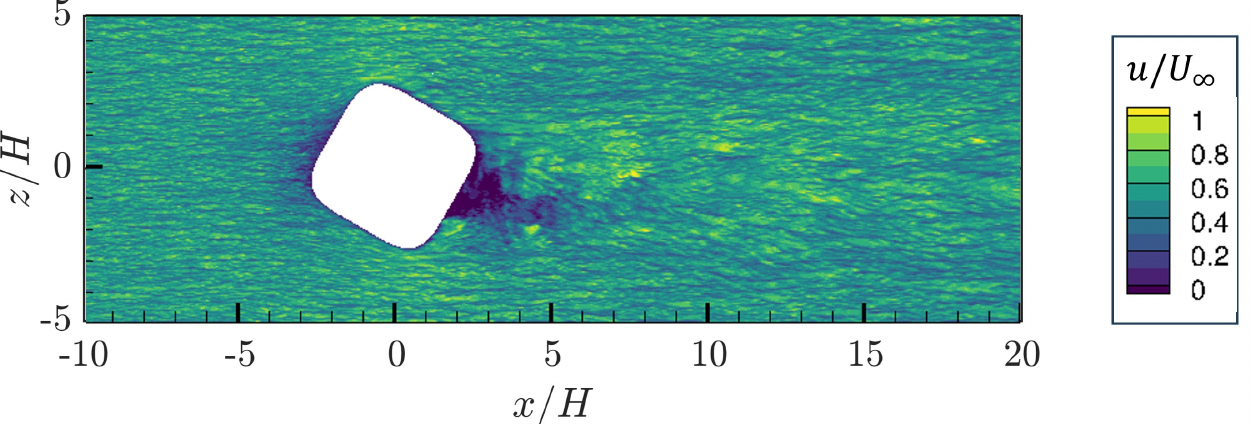}}
  \caption{Instantaneous velocity contours at $y/H = 0.02$ for flow past the BeVERLI hill.}
\label{fig:inst_vel}
\end{figure}

Figure~\ref{fig:inst_vel} shows instantaneous streamwise velocity contours at a wall-parallel plane very close to the surface ($y/H \approx 0.02$). A pronounced velocity-deficit region appears immediately downstream of the hill crest, indicating wake formation and turbulent separation. Upstream, streamwise streaks typical of wall-bounded turbulence are evident, but they are disrupted by separation in the wake before re-emerging further downstream.

\begin{figure}
\centerline{\includegraphics[width=1.0\linewidth]{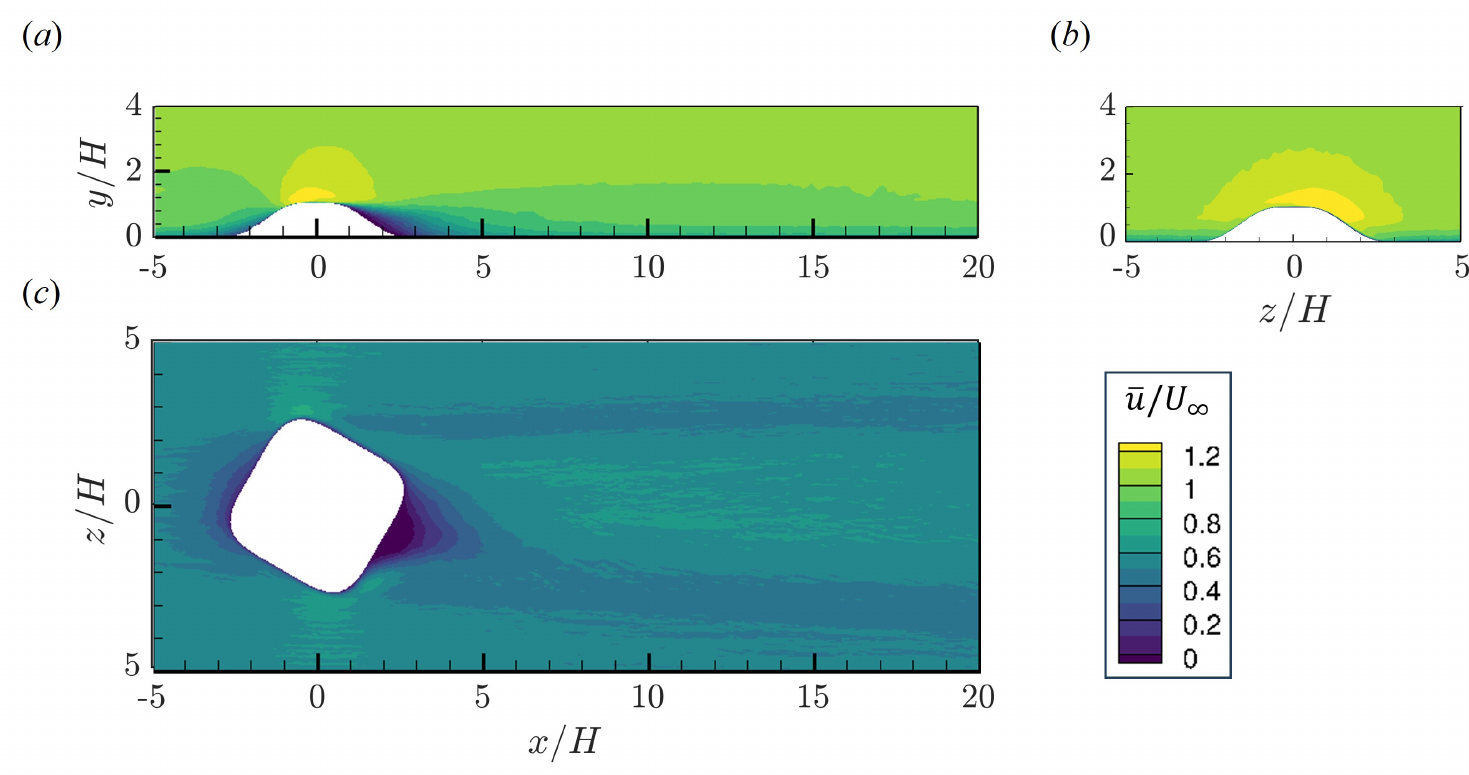}}
  \caption{Streamwise mean velocity contours normalized by $U_\infty$ at: (a) $z/H = 0$, (b) $x/H = 0$, and (c) $y/H \approx 0.02$.}
\label{fig:bhill_mean_vel}
\end{figure}

Streamwise mean velocity contours, normalized by the inlet freestream velocity, are shown in figure~\ref{fig:bhill_mean_vel}. Over the hill crest, the convex windward surface accelerates the flow, producing a favorable pressure gradient that stabilizes the boundary layer. Immediately downstream, the concave lee side induces an adverse pressure gradient, leading to a velocity deficit and separation region, as seen in figures~\ref{fig:bhill_mean_vel}(a) and \ref{fig:bhill_mean_vel}(c). Here, mean velocities drop below $U/U_\infty < 0.4$, marking the recirculation bubble behind the hill. Figure~\ref{fig:bhill_mean_vel}(b) illustrates the lateral spreading of the boundary layer.
The sharp transition from high to low velocity, coinciding with the change in wall curvature in figure~\ref{fig:bhill_mean_vel}(c), highlights the strong shear region associated with adverse pressure gradients. Further downstream, the boundary layer recovers and reattaches near the wall.

\begin{figure}
\centerline{\includegraphics[width=0.9\linewidth]{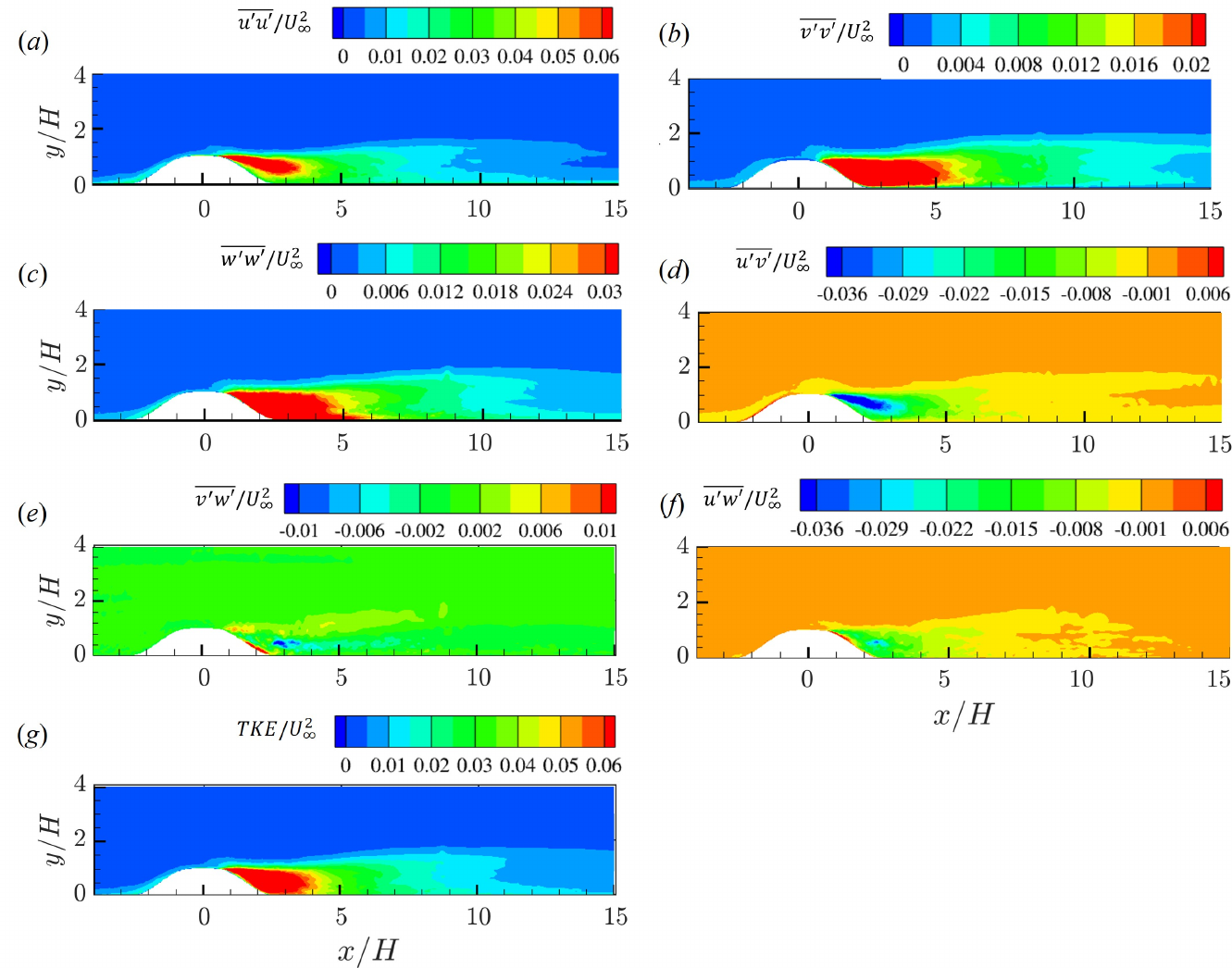}}
  \caption{Reynolds stresses (a–f) and turbulent kinetic energy (g) normalized by $U_\infty^2$ at the mid-plane $z/H=0$.}
\label{fig:bhill_mean_restresses}
\end{figure}

Figure~\ref{fig:bhill_mean_restresses} shows Reynolds stresses and turbulent kinetic energy (TKE) in the mid-plane ($z/H = 0$). Each component reflects the flow features identified in figure~\ref{fig:bhill_mean_vel}. The streamwise normal stress $\overline{u^\prime u^\prime}$ (figure~\ref{fig:bhill_mean_restresses}a) peaks downstream of the crest within the separated wake, reflecting strong fluctuations induced by separation and recirculation, and decays gradually downstream as turbulence dissipates. Similar patterns are observed for the normal (figure~\ref{fig:bhill_mean_restresses}b) and spanwise (figure~\ref{fig:bhill_mean_restresses}c) normal stresses, though at smaller magnitudes, reflecting anisotropy of the turbulence. The boundary layer thins approaching the hill, consistent with the growth indicated by these stresses.  
The Reynolds shear stress $\overline{u^\prime v^\prime}$ (figure~\ref{fig:bhill_mean_restresses}d) is negative, showing downward turbulent momentum transport from high-speed outer layers toward the near-wall region in the separated wake. The spanwise shear stresses $\overline{v^\prime w^\prime}$ (figure~\ref{fig:bhill_mean_restresses}e) and $\overline{u^\prime w^\prime}$ (figure~\ref{fig:bhill_mean_restresses}f), which vanish in purely two-dimensional flows, are clearly nonzero, highlighting the asymmetry, anisotropy, and lateral transport characteristic of this 3D geometry. 
Finally, the TKE distribution (figure~\ref{fig:bhill_mean_restresses}g) peaks in the wake downstream of the crest, coinciding with regions of intense Reynolds stresses. 

\subsubsection{Conventional model validation study}

\begin{figure}
\centerline{\includegraphics[width=0.8\linewidth]{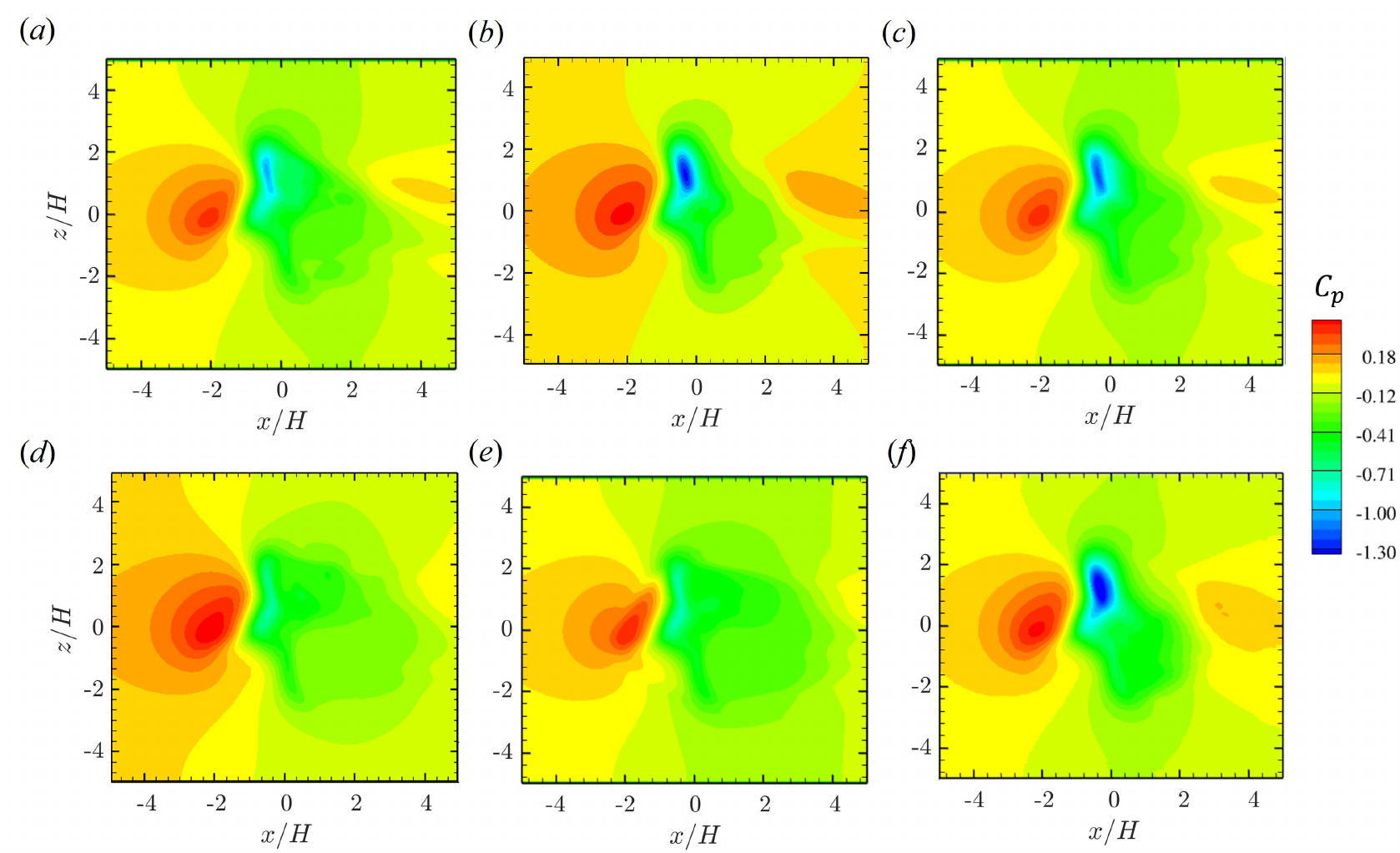}}
  \caption{Pressure coefficient ($C_p$) over the hill surface from: (a) $k-\omega$ SST, (b) SA, \\ (c) $k-\epsilon$, (d) $v^2$--$f$, (e) SSG–LRR FRSM, and (f) WRLES.}
\label{fig:bhill_cp}
\end{figure}

Conventional model validation typically focuses on engineering-relevant quantities such as the pressure coefficient and the skin-friction coefficient. Figure~\ref{fig:bhill_cp} compares the pressure-coefficient distribution over the hill surface predicted by different RANS turbulence models against WRLES. All models reproduce the general pressure topology: a high-pressure region upstream of the hill, a distinct low-pressure region near the crest, and pressure recovery downstream. However, the $k$--$\epsilon$, $k-\omega$ SST, $v^2-f$ and SSG–LRR FRSM models fail to capture the pressure minimum at the crest. Moreover, the downstream pressure surge evident in WRLES is absent in the $v^2-f$ and SSG-LRR FRSM results, suggesting an overprediction of separation-bubble size. The SA model better predicts the minimum but overestimates pressure on the sides of the hill and during recovery.

\begin{figure}
\centerline{\includegraphics[width=0.8\linewidth]{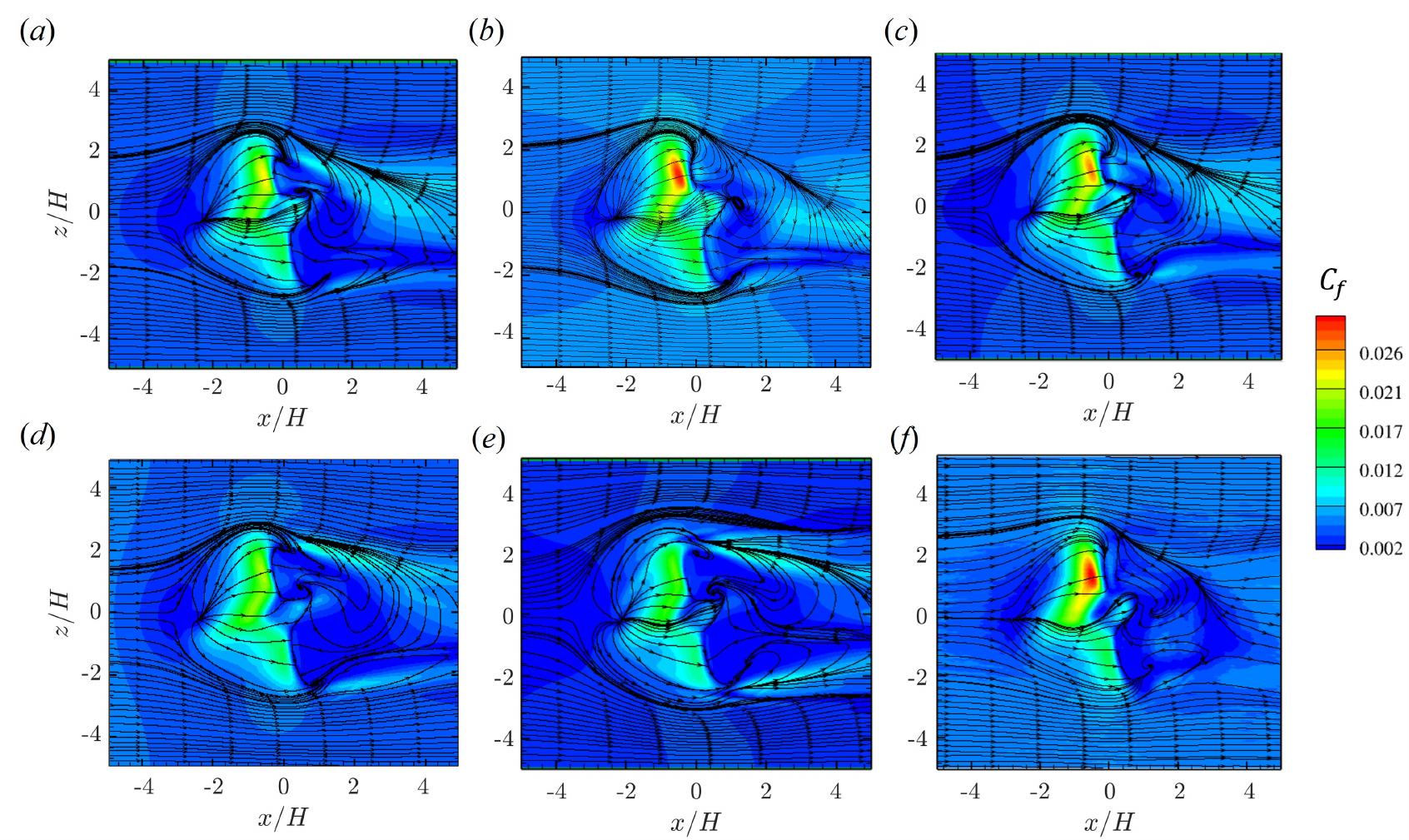}}
  \caption{Skin-friction coefficient ($C_f$) over the hill surface from: (a) $k-\omega$ SST, (b) SA, (c) $k-\epsilon$, (d) $v^2$--$f$, (e) SSG–LRR FRSM, and (f) WRLES.}
\label{fig:bhill_cf}
\end{figure}

Figure~\ref{fig:bhill_cf} presents the corresponding skin-friction coefficient distributions. The WRLES reference (figure~\ref{fig:bhill_cf}e) shows low $C_f$ just upstream of the hill, elevated $C_f$ on the windward face, a sharp drop on the lee side, and eventual recovery downstream. The RANS models reproduce this overall pattern, but with important discrepancies. The $k$--$\epsilon$ and SA models agree reasonably well near the crest but overpredict $C_f$ downstream. In contrast, the $k-\omega$ SST, $v^2-f$ and SSG-LRR FRSM models underpredict the peak $C_f$ near the reattachment region. All models overpredict $C_f$ throughout most of the wake. Surface streamlines of mean $(\overline{u},\overline{w})$ also agree qualitatively with WRLES: a horseshoe vortex upstream, curvature-driven deflection on the hill surface, and recirculation downstream. However, deviations emerge in separation patterns, which curve outward more strongly in some models as they extend downstream.

From the conventional CFD perspective, validation would stop at figures~\ref{fig:bhill_cp} and \ref{fig:bhill_cf}, where $C_p$ and $C_f$ distributions (and associated surface streamlines) are compared to assess model performance against WRLES. While such comparisons are necessary, they are insufficient: they provide little insight into which aspects of the underlying flow physics are misrepresented by the models.


\subsection{Model Error Diagnostics for BeVERLI Hill} \label{sec:4.3}

In this subsection, we apply the error diagnostics introduced in \S\ref{sec:2} to the BeVERLI hill. The analysis focuses on the flat-plate region rather than on the hill itself.
In complex three-dimensional geometries, the surface curvature complicates the definition of key quantities required for the integral formulation, including the boundary-layer thickness, the freestream velocity, and the appropriate wall-normal integration limits. 
In principle, the AMI decomposition could be reformulated in a fully curvilinear coordinate system aligned with the local surface geometry. 
Such an approach has been demonstrated for two-dimensional bumps and airfoils \citep{kianfar2025moment}. 
However, extending this framework to a fully three-dimensional curved surface is beyond the scope of the present study.
For clarity, we present results for the $k-\omega$ SST model; results for the other RANS models are provided in the supplementary materials.

\begin{figure}
\centerline{\includegraphics[width=0.9\linewidth]{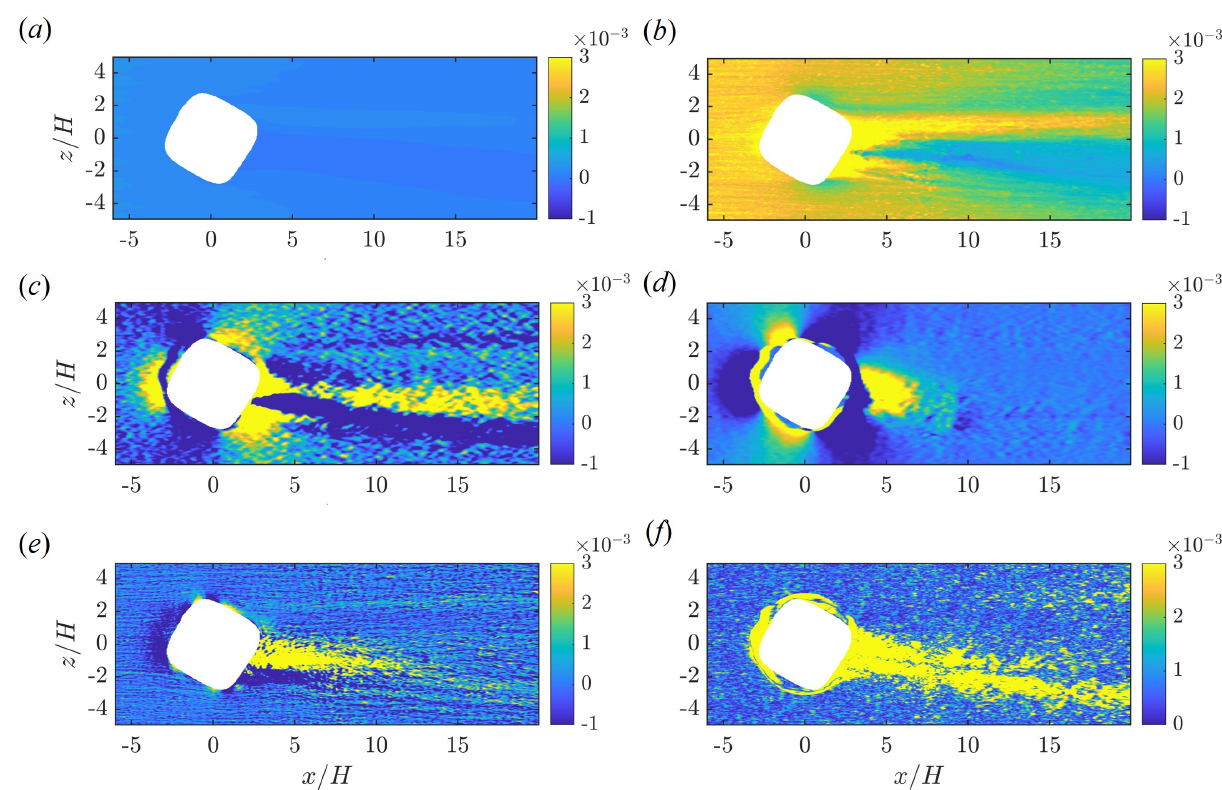}}
  \caption{AMI contributions from WRLES: (a) Laminar friction, (b) Turbulent torque, (c) Total mean flux, (d) Freestream pressure gradient, (e) Departure from BL approximation, and (f) Residual $(|C_f/2 - \text{RHS}|)$.}
\label{fig:bhill_les_ami}
\end{figure}

We first present the AMI decomposition results for the WRLES solutions. Figure~\ref{fig:bhill_les_ami} shows the contributions from viscosity, turbulent torque, mean-flow convection, freestream pressure gradients, departures from the boundary-layer approximation, and the residual.
It can be observed that these decompositions exhibit strong spatial variations.
For visual consistency across panels, a fixed color scale is used, with values outside the specified range clipped to the extrema.
The viscous contribution in figure~\ref{fig:bhill_les_ami}(a) is small throughout the domain.
The turbulent torque component in figure~\ref{fig:bhill_les_ami}(b) shows higher values in the near wake region, gradually reducing farther downstream in an asymmetric manner.
Note that this turbulence term stems from the unweighted integral of the Reynolds shear stress.
Consequently, the outer layer plays a major role in determining the component \citep{kianfar2025moment}.
The variations in $\overline{u^\prime v^\prime}$ in figure~\ref{fig:bhill_mean_restresses}(d) illustrate this effect, leading the turbulent torque contributions at the mid-plane ($z/H = 0$) to peak in the near-wake region and then drop off farther downstream.
The mean-flux term in figure~\ref{fig:bhill_les_ami}(c) absorbs some of the excess torque into thickening of the boundary-layer flow by growing its
associated angular momentum. 
The term exhibits spatial variations near the hill and in the wake region.
This is in contrast to the flat-plate turbulent boundary layer case in figure~\ref{fig:bl_ami}, where the mean-flux term uniformly attenuates $C_f$.
The pressure-gradient contribution in figure~\ref{fig:bhill_les_ami}(d) shows sign changes as the turbulent boundary layer flow approaches the hill: it becomes strongly negative in the adverse-pressure-gradient region in the wake and switches sign again as the flow re-accelerates after separation, contributing positively to $C_f$ in the recovery region.
A similar sign-alternating behavior in the pressure-gradient AMI term is observed qualitatively in the 2D Gaussian bump case \citep{kianfar2025moment}.
The departure from boundary layer approximation in figure~\ref{fig:bhill_les_ami}(e) is most pronounced in the separated flow region, as would be expected. 
The residual term, shown in figure~\ref{fig:bhill_les_ami}(f), represents the difference between the sum of these contributions (the RHS of equation~\eqref{eq:AMI}) and $C_f/2$.
This residual appears small but still non-negligible across most of the domain.
As discussed in \cite{elnahhas2022enhancement} and further quantified in \S\ref{sec:2.1}, these residuals arise from two main sources: inaccuracies in identifying the freestream velocity $U_{\infty}(x)$ and non-zero fluctuations of the pressure gradient outside the boundary layer due to insufficient time averaging. The integral truncation limit, currently set at $1.5 \delta$, has also been found to influence the magnitude of these residuals.
Here, the boundary layer thickness $\delta$ denotes the distance from the wall to the point where the streamwise mean velocity reaches 95\% of the irrotational velocity $U_{io}$.
Similar AMI analyses can be performed to evaluate the integrals on solutions obtained from RANS models. However, while such analysis provides insights into the various flow physics that contribute to the skin friction, it does not offer insights into the modeling error.

\begin{figure}
\centerline{\includegraphics[width=0.92\linewidth]{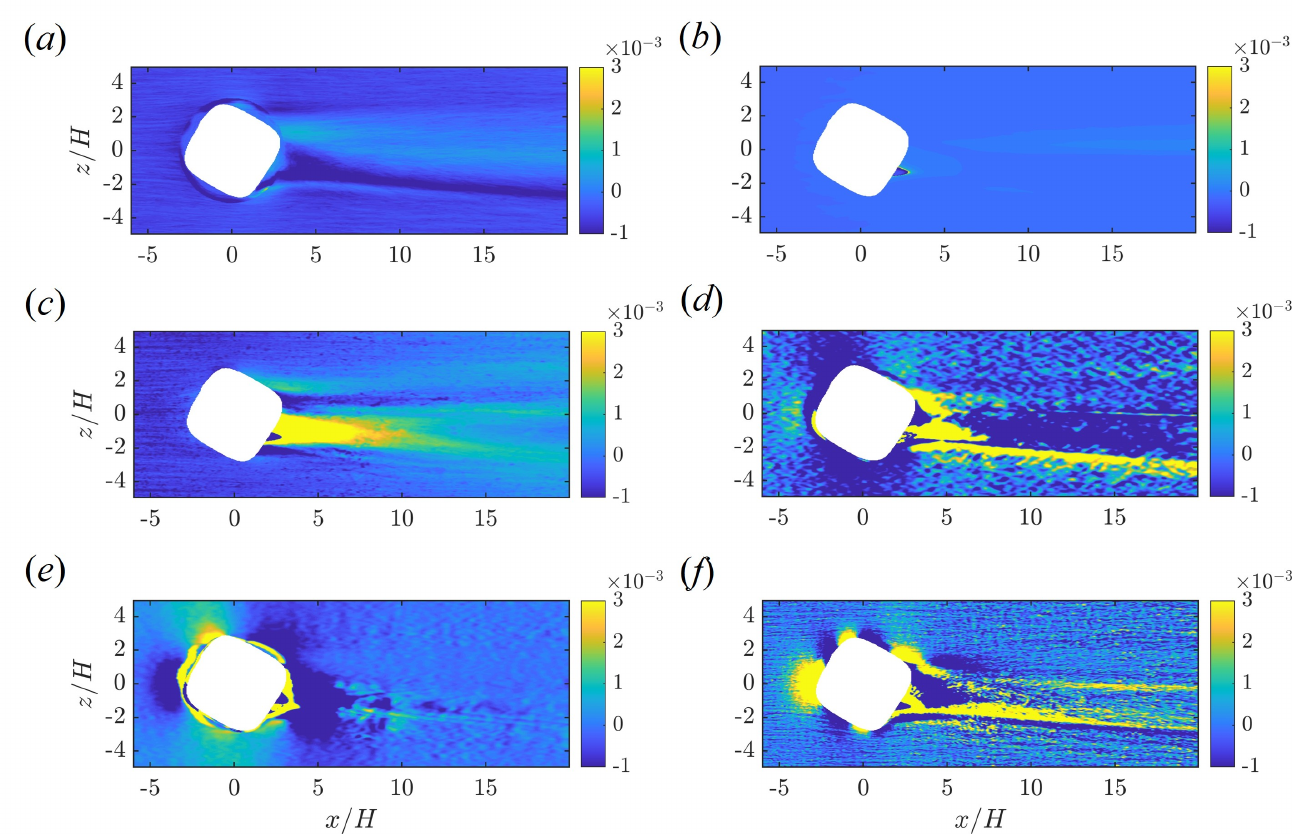}}
  \caption{AMI based error analysis as per \eqref{eq:error_analysis} for $k-\omega$ SST model predictions showing differences in: (a) $C_f/2$, (b) Laminar friction, (c) Turbulent torque, (d) Total mean-flux, (e) Free-stream pressure gradient and (f) Departure from BL approximation.}
\label{fig:bhill_sst_error}
\end{figure}

Next, we demonstrate the error diagnostic framework in equation \eqref{eq:error_analysis} for the $k-\omega$ SST model.
Figure \ref{fig:bhill_sst_error}(a) depicts the overall error in $C_f/2$ present in the model when WRLES is taken as the reference data.
At first glance, the model can be observed to underpredict the skin-friction in the immediate wake and overpredict on the $+z$ side of the wake.
This is also apparent when comparing figures \ref{fig:bhill_cf}(e) and \ref{fig:bhill_cf}(a), which depict $C_f$ for WRLES and $k-\omega$ SST respectively.
While the model overpredicts turbulent torque term (figure \ref{fig:bhill_sst_error}c), yet this is over-suppressed by underpredicted freestream pressure gradient contributions (figure \ref{fig:bhill_sst_error}e) in the near-wake and by excess mean-flux attenuation (figure \ref{fig:bhill_sst_error}d) farther downstream.
There is no major deviation in the viscous component in figure \ref{fig:bhill_sst_error}(b).
The departure-from-BL term \ref{fig:bhill_sst_error}(f)  mirrors the mean-flux streaks with alternating signs near the hill, flagging 3-D curvature/separation influences that steady RANS cannot fully capture.
In short, the framework reveals that the $k-\omega$ SST model's errors in the wake stem from misallocation of the skin-friction budget—too much turbulent transport, coupled with an under-response to adverse pressure gradients and an over-suppressed boundary-layer growth downstream.
Analogous term-wise error maps for SA, $k-\epsilon$ and SSG-LRR FRSM models have been reported in the supplementary material for brevity.

\begin{figure}
\centerline{\includegraphics[width=1.0\linewidth]{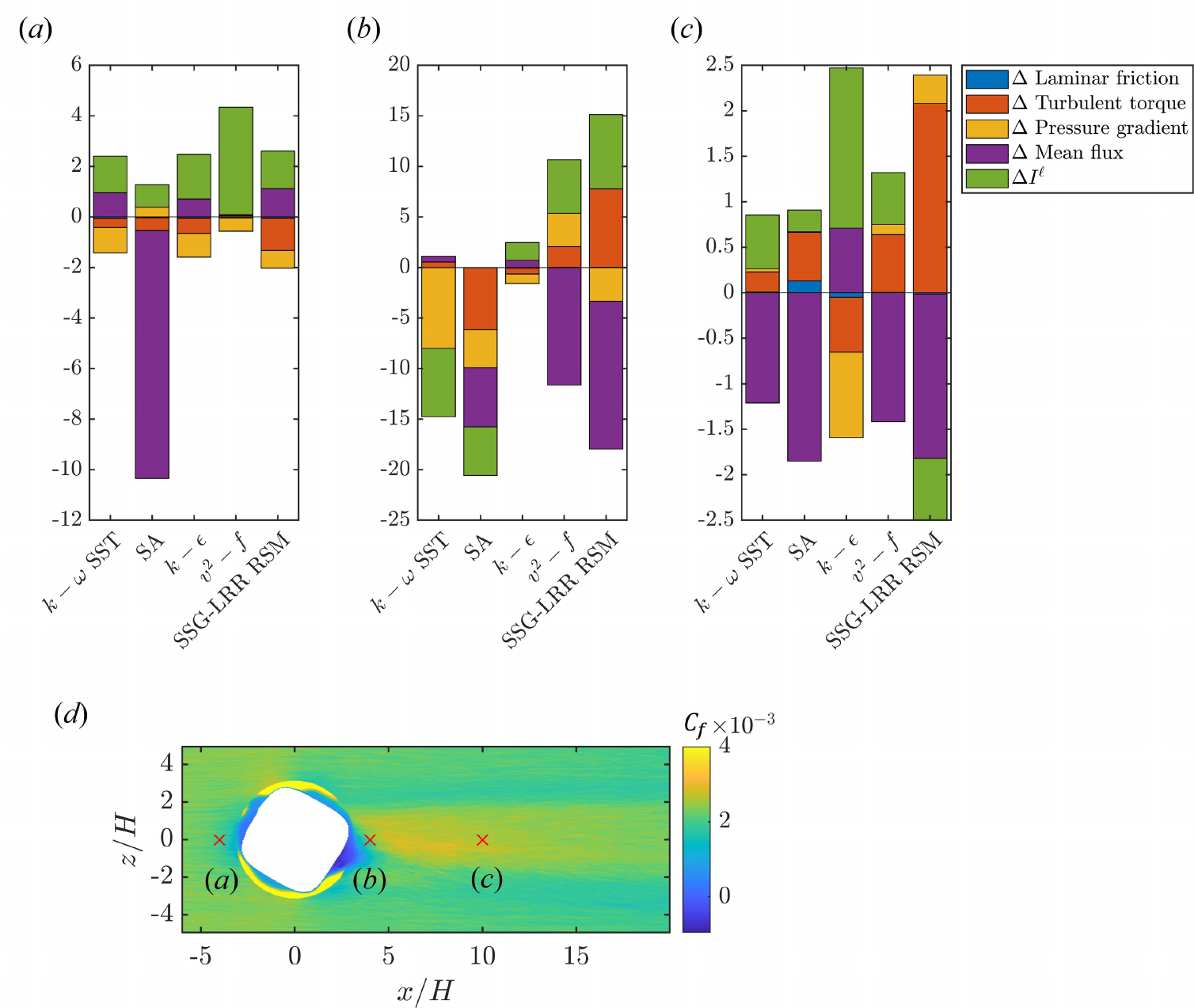}}
  \caption{BeVERLI hill $C_f$ decomposition errors for different models at the indicated locations: (a) $x/H = -4$, (b) $x/H = 4$, and (c) $x/H = 10$. The error in each contribution is normalized by $C_{f,\text{WRLES}}/2$.}
\label{fig:bhill_Cf_decomp_percent}
\end{figure}

We conduct a more detailed model error analysis at three representative locations.  
Figure~\ref{fig:bhill_Cf_decomp_percent} depicts errors in $C_f$ contributions, extracted at three different locations as marked in \ref{fig:bhill_Cf_decomp_percent}(d) for the five turbulence models.  
These locations are placed on the center-span ($z/H=0$) and correspond to different streamwise positions, $x/H = -4$, $x/H = 4$, and $x/H = 10$, respectively.  
At $x/H = -4$, near the windward face of the hill, most models exhibit component-wise errors of O(1-4) with varying under- or overpredicted components, when normalized by the local WRLES $C_f/2$. 
The SA model appears an outlier, showing a much larger discrepancy dominated by the mean-flux contribution ($\approx10$x WRLES $C_f/2$).
The departure from boundary-layer approximation component is the leading source of error in the $k$--$\omega$ SST, $k$--$\epsilon$, $v^2-f$ and SSG-LRR RSM models. 
This could be intuitively explained by the weak adverse pressure gradient induced by the hill upstream that modifies the local streamwise development and introduces three-dimensional effects not captured by classical boundary-layer assumptions. 
This is evident in where the $C_p$ contours (figure \ref{fig:bhill_cp}) and corresponding $C_f$ distributions (figure \ref{fig:bhill_cf}) at $x/H=-4$ being affected by the approaching geometry.
For the SA model, the mean-flux term is the dominant contributor and is underpredicted at this location.  
Unlike the flat-plate case, the errors in turbulent torque are minimal when compared to other components except for the SSG-LRR RSM case.
At location $x/H = 4$, which is just downstream of the hill, the errors are greatest in the SSG-LRR FRSM model, with variations as high as fifteen times $C_f/2$ in the mean flux term.  
Interestingly, while the individual error components show large underpredictions in the mean-flux component and overpredictions in the turbulent torque component, they also tend to effectively cancel each other to produce reasonable $C_f$ values.  
Farther downstream at $x/H = 10$, this cancellation behavior between mean-flux and turbulent torque is observed in varying degrees for all the models.
The $k-\epsilon$ model shows a strong under prediction in freestream pressure-gradient contribution, balanced largely by over prediction in departure from BL approximation term $\Delta I^\ell$.
In all models, the individual components remain comparable to or somewhat larger than $C_{f,\text{WRLES}}/2$.
The error in the viscous term is negligible for all models at all three locations.  
It is also important to note that, unlike the flat-plate $C_f$ decompositions shown in figure~\ref{fig:bl_Cf_decomp_percent}, the error components in the BeVERLI hill case do not cancel as effectively. This reduced compensation is particularly evident in the favourable pressure-gradient region at $x/H=-4$ (figure~\ref{fig:bhill_Cf_decomp_percent}(a)) and in the wake region at $x/H=4$ (figure~\ref{fig:bhill_Cf_decomp_percent}(b)), especially for the $k$--$\omega$ SST, SA, and $k$--$\epsilon$ models.
This reduced compensation reflects the increased sensitivity of complex three-dimensional, separated flow to imbalances among the AMI contributions.

\begin{figure}
\centerline{\includegraphics[width=0.9\linewidth]{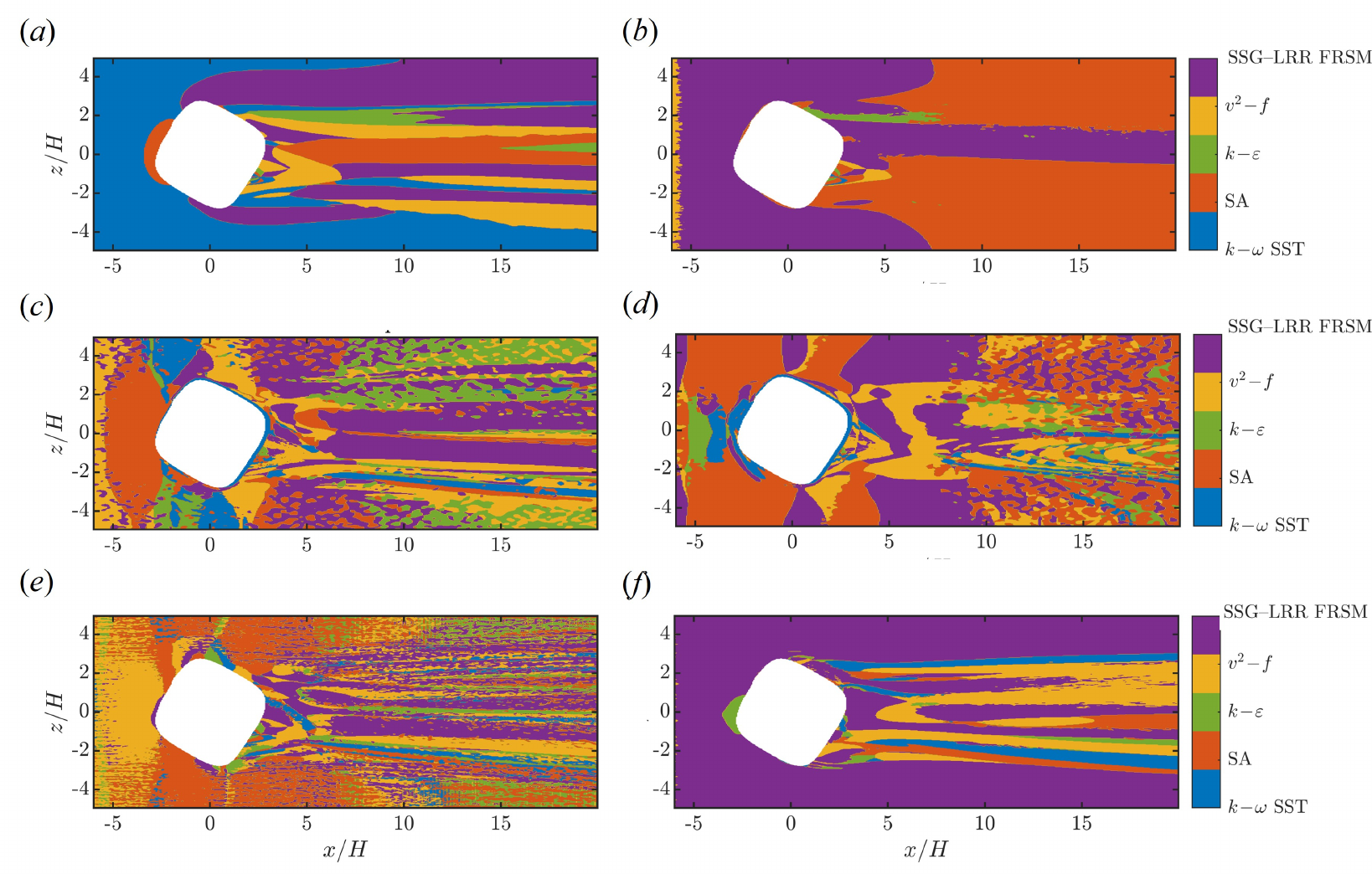}}
  \caption{Worst offender plot indicating which model yields the highest error in each AMI component: (a) Laminar friction, (b) Turbulent torque, (c) Total mean-flux, (d) Free-stream pressure gradient, (e) Departure from BL approximation, and (f) $C_f$.}
\label{fig:wo}
\end{figure}

For engineering practices, one often needs to pick a model.  
This will require knowledge of the relative performance of the existing models.  
Figure~\ref{fig:wo} depicts the regions based on which model performs the worst and has the largest error relative to WRLES in magnitude in its $C_f$ decomposition.  
Such comparative analysis helps understand the behavior of RANS models in complex hill-type flows and isolate model deficiencies in predicting TBLs with pressure gradients.  
Figure~\ref{fig:wo}(a) depicts which model performs worst in predicting the minor viscous component, with the $k$--$\omega$ SST model exhibiting the most errors upstream, while other models perform worse downstream of the hill.  
SA and SSG-LRR FRSM models perform worst in predicting turbulent torque in figure~\ref{fig:wo}(b), which is a leading contributor to $C_f$.  
In figure~\ref{fig:wo}(c), the mean-flux term is significantly overpredicted by the SA and $v^2-f$ models compared to the others, particularly upstream.  
In figure~\ref{fig:wo}(d), the eddy-viscosity models—SA, $k$--$\omega$ SST, and $k$--$\epsilon$—perform poorly in predicting the pressure-gradient effects component, whereas the SSG-LRR FRSM and $v^2-f$ perform poorly in the wake region.  
The worst-offender plot in figure~\ref{fig:wo}(e) shows models mispredicting 3D, unsteady, and flow-separation effects, which are generally neglected in boundary-layer theory—confirming that no steady RANS model can fully capture them.  
Figure~\ref{fig:wo}(f) encapsulates the cumulative effect on the skin-friction coefficient, where the SSG-LRR FRSM model is the most inaccurate across most regions, followed by the $v^2-f$ model in the wake.

\section{Conclusion} \label{sec:5}

This study presents a practical methodology for diagnosing missing physics in turbulence models by decomposing skin friction ($C_f$) into physically interpretable components. Building on the AMI equation proposed by \citet{elnahhas2022enhancement}, and extending it to include spanwise transport, we derive a $C_f$ decomposition applicable to three-dimensional turbulent boundary layers. This decomposition isolates contributions from viscosity, turbulent torque, mean-flux growth, pressure-gradient effects, and departures from boundary-layer assumptions. Using AMI terms derived from DNS or WRLES as high-fidelity baselines, we construct detailed error budgets for standard RANS closures. The resulting framework goes beyond conventional verification and validation, revealing how seemingly accurate $C_f$ predictions may arise from compensating errors.
More broadly, the results demonstrate that error cancellation is a pervasive feature in RANS predictions of quantities of interest. 
Substantial local modeling errors can propagate through the flow solution in such a way that their net impact on integral quantities is reduced. 
While this behavior may be reassuring and consistent with the traditional objective of turbulence modeling — namely, the accurate prediction of engineering quantities— it also highlights the risk that accurate $C_f$ predictions may mask significant misrepresentation of the underlying flow physics.
For instance, if errors cancel in a canonical case, then it is also likely that these errors might compound for a more complex case.
The present framework therefore complements conventional validation by exposing how different processes contribute to the predicted aggregate skin friction.

The methodology was applied to both a zero-pressure-gradient flat-plate turbulent boundary layer and a more complex turbulent boundary layer flow over a three-dimensional bump. Five turbulence models were evaluated: $k$--$\omega$ SST, SA, $k$--$\epsilon$, $v^2-f$ and SSG-LRR FRSM. In the flat-plate case, all models reproduced reasonably accurate $C_f$ values, primarily through cancellation of large but opposing errors. These typically involved overprediction of the turbulent torque term, balanced by underprediction of the mean-flux term, with individual errors reaching up to 25\% of $C_f$. 
For the three-dimensional case, WRLES was performed over the BeVERLI hill geometry, inclined at $30^\circ$ to the streamwise direction. This served as the reference. 
As with the flat-plate case, the turbulent torque was found to be the dominant contributor to $C_f$ across much of the domain. However, in the hill wake region, the contributions from mean-flux growth, pressure-gradient effects, and boundary layer approximation departure terms became substantial, highlighting the influence of three-dimensionality and flow separation. Spatially resolved error maps for the RANS models revealed where these effects were missed, offering insight into the physical limitations of each closure. In particular, the SSG-LRR FRSM model exhibited errors in $C_f$ contributions exceeding twenty times $C_f/2$ at certain locations, underscoring the extreme nature of compensating errors in complex flows.

The present framework opens several avenues for future work. 
First, the proposed methodology is complementary to turbulence modeling and data-driven approaches. 
By comparing DNS/LES-based and RANS-predicted $C_f$ decompositions, a resulting physics-based error map can be used to interpret the outcome of model and evaluate whether proposed closure changes improve the right mechanisms and avoid improvements in $C_f$ that arise purely from error cancellation. 
For example, a model may overpredict the turbulent-torque contribution while underpredicting the mean-flux contribution, leading to an apparently accurate $C_f$.
In such cases, correcting only one of these deficiencies in isolation may actually worsen the overall prediction by removing the compensating effect. 
Conversely, in separated regions of the hill flow, errors in multiple mechanisms may compound rather than cancel, leading to large deviations in $C_f$. 
By identifying whether inaccuracies are compensating or compounding, the framework helps prioritize closure modifications in a physically consistent manner.
Beyond this error-balancing perspective, the diagnostics may also provide other insights. 
For instance, while the present work does not isolate the dissipation rate ($\epsilon$) equation errors explicitly, systematic misrepresentation of mean-flux growth indirectly reflects deficiencies in predicted
boundary-layer thickness evolution, which depends on production–dissipation balance \cite{zpgtbl}.
Hence, the diagnostics provide indirect but mechanistically interpretable feedback on dissipation modeling.
Second, by associating specific error components with modeling assumptions—such as the Boussinesq hypothesis—the framework provides a tool for critically assessing and refining turbulence modeling paradigms, thereby advancing our fundamental understanding of turbulence closures.
For instance, the turbulent torque term is directly related to the modeled Reynolds shear stress.
Persistent overprediction of turbulent torque (as observed in the flat-plate case) indicates systematic bias in eddy-viscosity magnitude or stress–strain alignment. 
This points directly to limitations of linear eddy-viscosity assumptions \citep{pope1975more} and anisotropy modeling \citep{speziale1990analytical}. 
Although Reynolds-stress anisotropy is not explicitly parameterized in the analysis, it is implicitly reflected in several AMI terms. 
For instance, anisotropy manifests in the form of turbulent-torque contribution through misprediction of the Reynolds shear stress and its wall-normal distribution.
Errors in the mean-flux and departure from BL approximation terms also reflect deficiencies in capturing cross-stream and spanwise transport — key to modeling anisotropy in 3D flows. 
Lastly, the current decomposition is limited to $C_f$, and extending the analysis to pressure coefficient will also be relevant to fluids engineering. 

\backsection[Supplementary data]{\label{SupMat} Additional figures showcasing AMI results, error analyses and noise analyses for other RANS models for both the flat-plate and B-hill cases are attached as supplementary material.}


\backsection[Funding]{Yang acknowledges AFOSR Grant No. FA9550-23-1-0272 with Dr. Gregg Abate as the technical monitor.
Kunz acknowledges ONR Grant N00014-24-1-2170 with Peter Chang and Julie Young as technical monitors.
Computational resources were provided through Penn State ROAR and a United States Department of Defense (DoD) Frontier project of the High Performance Computing Modernization Program (HPCMP).
}

\backsection[Declaration of interests]{The authors report no conflict of interest.}

\backsection[AI statement]{During the preparation of this work, the authors used ChatGPT in order to make grammatical corrections. After using this tool/service, the authors reviewed and edited the content as needed and take full responsibility for the content of the publication.}


\backsection[Author ORCIDs]{Shyam S. Nair, https://orcid.org/0009-0009-9284-3075; Vishal A. Wadhai, https://orcid.org/0000-0003-0521-9784; Robert F. Kunz, https://orcid.org/0000-0001-9504-1945; Xiang I. A. Yang, https://orcid.org/0000-0003-4940-5976}

\appendix
\section{Alternative groupings of AMI contributions}\label{appA}

In equation \eqref{eq:AMI}, the term $I_\ell$ collects contributions that represent departures
from the classical boundary-layer approximation, including effects of unsteadiness,
spanwise freestream pressure gradients, and additional three-dimensional mean-flow
terms arising from equation \eqref{eq:Ix}. 
In the present work, these terms are grouped
into a single departure from boundary-layer approximation contribution for
clarity of physical interpretation.

However, alternative groupings are possible. In particular, the first boxed term
in equation \eqref{eq:Ix}, $(W-\bar w)\frac{\partial U}{\partial z}$, represents the effect of spanwise freestream pressure gradients and could
alternatively be grouped together with the streamwise pressure-gradient
contribution in equation \eqref{eq:Ix}. 
Carrying this change through the AMI derivation
would modify the decomposition in \eqref{eq:Ix} to \eqref{eq:AMI2} by transferring part of $I^\ell$ into the pressure-gradient term, without altering the exact identity for $C_f$.

\begin{align} \label{eq:AMI2}
    \frac{C_f}{2} &= \frac{1}{Re_{\ell}} 
    + \int_0^{\infty} \frac{-\overline{u^{\prime} v^{\prime}}}{U_{io}^2 \ell} \, \mathrm{d}y 
    + \frac{\partial \theta_{\ell x}}{\partial x} 
    - \frac{\theta_x - \theta_{\ell_x}}{\ell} \frac{\mathrm{d}\ell}{\mathrm{d}x} \nonumber 
     + \mathstrut\displaystyle\frac{\partial \theta_{\ell z}}{\partial z}
     - \mathstrut\displaystyle\frac{\theta_z - \theta_{\ell_z}}{\ell}\frac{\mathrm{d}\ell}{\mathrm{d}z}
        \\ & \quad
    + \frac{\theta_v}{\ell} 
    + \frac{\delta_{\ell_x}^* + 2 \theta_{\ell_x}}{U_{io}} \frac{\partial U}{\partial x} 
    + \mathstrut\displaystyle\frac{\delta_{\ell_z}^*+ 2 \theta_{\ell_z}}{U_{io}}\frac{\partial U}{\partial z}  
    + \mathcal{I}^{\ell}_*.
\end{align}

where

\begin{equation}\label{eq:Ix2}
\begin{split}
I_x^* \equiv \frac{\partial\left(U-\bar{u}\right)}{\partial t}
+ \left(V-\bar{v}\right)\frac{\partial U}{\partial y}
+ \frac{1}{\rho} \frac{\partial\left(P-\bar{p}\right)}{\partial x}\\
- \nu\left[ \frac{\partial^2 (U-\bar{u})}{\partial x^2} +\frac{\partial^2 U}{\partial y^2}
+ \mathstrut\displaystyle\frac{\partial^2 (U-\bar{u})}{\partial z^2} \right]
- \frac{\partial \overline{u^{\prime 2}}}{\partial x}
- \mathstrut\displaystyle\frac{\partial \overline{u^{\prime}w^{\prime}}}{\partial z}
\end{split}
\end{equation}

and 

\begin{equation}
\mathcal{I}^{\ell}_{*} \equiv \int_0^{\infty}\left(1-\frac{y}{\ell}\right) \frac{I_x^*}{U_{io}^2} \mathrm{~d} y, ~~      \delta_{\ell_z}^* \equiv \int_{0}^{\infty} \left(1 - \frac{y}{\ell}\right) \left(\frac{W - \bar{w}}{U_{io}}\right) \, \mathrm{d}y
\end{equation}

\begin{figure}
\centerline{\includegraphics[width=1.0\linewidth]{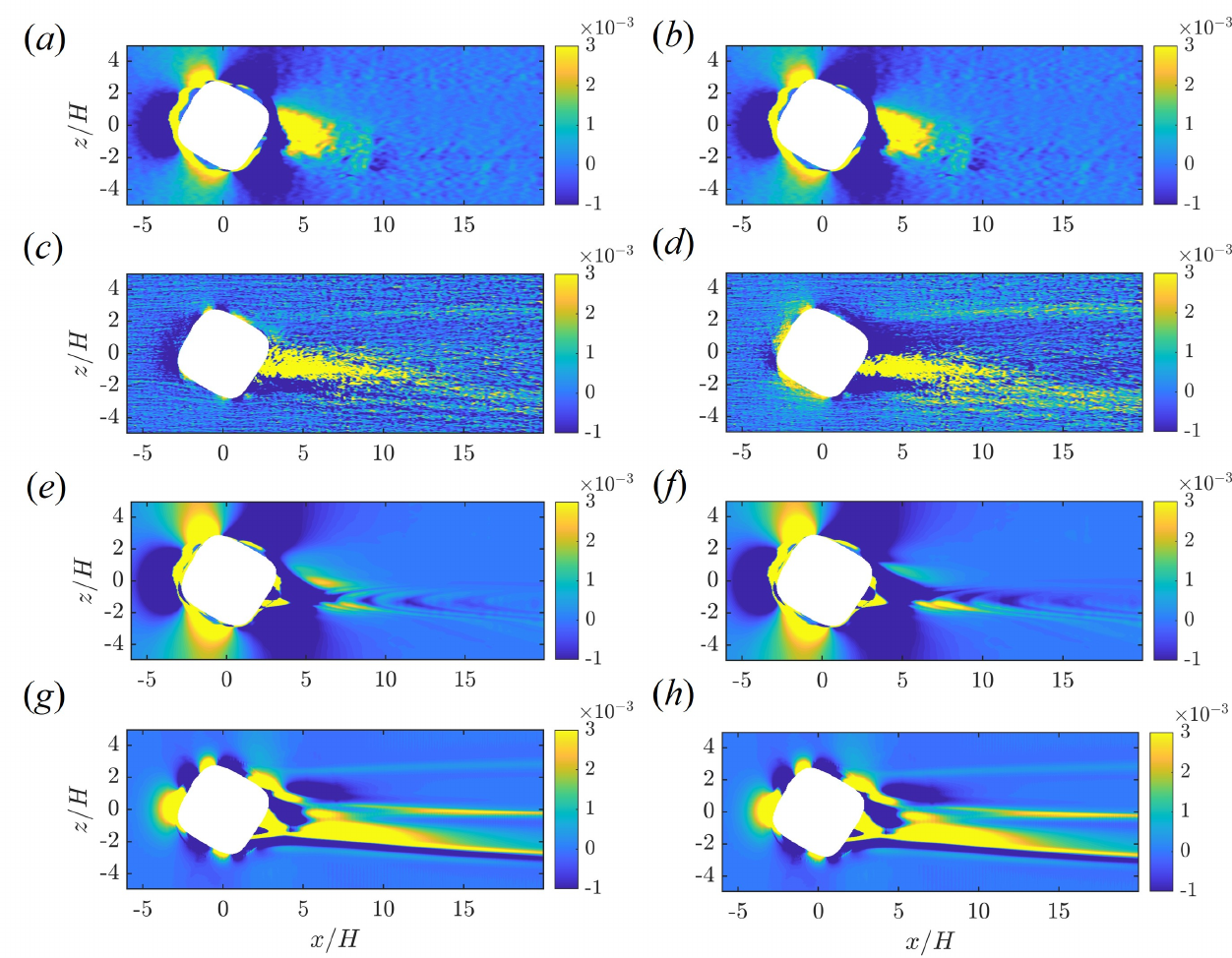}}
\caption{AMI contributions under alternative groupings of the spanwise freestream pressure gradient terms: 
(a,c,e,g) Original grouping as per \eqref{eq:AMI} and (b,d,f,h) alternate grouping as per \eqref{eq:AMI2}. 
(a,b,e,f) show the pressure gradient contribution and (c,d,g,h) the departure from BL term. (a–d) WRLES; (e–h) SST.}
\label{fig:appendix_fig}
\end{figure}

To assess the impact of such regrouping, we recomputed the AMI decomposition for such an alternative classification.
Figure \ref{fig:appendix_fig} compares the resulting free-stream pressure gradient and departure from BL approximation terms that arise from regrouping in for both WRLES and $k-\omega$ SST solutions.
Some differences between the original and revised groupings can be observed in the departure from BL term, particularly in figures \ref{fig:appendix_fig}(c) and \ref{fig:appendix_fig}(d). 
These variations are primarily located in the FPG and APG regions near the hill, where spanwise effects are more pronounced. 
However, the overall spatial distribution and qualitative trends remain similar under both groupings.

Different groupings may lead to different component-wise agreement between RANS and WRLES predictions.
The grouping adopted in \eqref{eq:AMI} maintains a distinction between canonical boundary-layer mechanisms and the 3D hill case, where spanwise effects are grouped as additional contributions arising from three-dimensionality and departure from classical boundary-layer assumptions. 
At the same time, the alternative grouping in \ref{eq:AMI2} may be advantageous when comparing multiple complex geometries involving separation, where spanwise freestream effects can reasonably be interpreted alongside pressure-gradient contributions.
The choice of whether to interpret spanwise freestream gradient effects as part of the departure from boundary layer term or to combine them with the freestream pressure gradient contribution is, to some extent, a matter of classification. 
Both groupings are mathematically consistent within the AMI formulation.
The methodology presented in this work remains interpretable under either choice, provided that the physical content represented by each grouped mechanism is clearly defined.


\newpage
\bibliographystyle{jfm}
\bibliography{jfm}

\end{document}